\documentclass[sn-apa,iicol]{sn-jnl}
\usepackage{graphicx}%
\usepackage{multirow}%
\usepackage{amsmath,amssymb,amsfonts}%
\usepackage{amsthm}%
\usepackage{manyfoot}%
\usepackage{adjustbox}
\usepackage{dirtytalk}
\usepackage{bm}
\usepackage{stmaryrd}
\usepackage{braket}
\usepackage{tabularx,array}
\usepackage{pgfplots}
\usepackage{caption}
\usepackage{subcaption}
\usepackage{mathtools}
\usepackage{tikz}
\usepackage{lscape}
\usepackage{orcidlink}
\usepackage{bbm}
\usepackage{breqn}
\usepackage{longtable}
\usepackage{booktabs}
\usepackage{array}

\pgfplotsset{compat=1.18}

\usepackage{nomencl}
\makenomenclature

\begin{document}

\title[Generalization Error Bounds for QML]{Generalization Error Bound for Quantum Machine Learning in NISQ Era - A Survey}

\author{\fnm{Bikram} \sur{Khanal}\orcidlink{0000-0003-2292-520X}}\email{bikram\_khanal1@baylor.edu}
\author*{\fnm{Pablo} \sur{Rivas}\orcidlink{0000-0002-8690-0987}}\email{pablo\_rivas@baylor.edu}
\author{\fnm{Arun} \sur{Sanjel}\orcidlink{0009-0009-2453-8538}}\email{arun\_sanjel1@baylor.edu}
\equalcont{These authors contributed equally to this work.}
\author{\fnm{Korn} \sur{Sooksatra}\orcidlink{0000-0003-4521-2237}}\email{korn\_sooksatra1@baylor.edu}
\equalcont{These authors contributed equally to this work.}
\author{\fnm{Ernesto} \sur{Quevedo}\orcidlink{0000-0002-8938-2230}}\email{ernesto\_quevedo1@baylor.edu}
\author{\fnm{Alejandro} \sur{Rodriguez}\orcidlink{0000-0003-0095-4077}}\email{alejandro\_rodriguez4@baylor.edu}

\affil{\orgdiv{School of Engineering \& Computer Science}, \orgname{Baylor University}, \orgaddress{\street{One Bear Place}, \city{Waco}, \postcode{76798}, \state{TX}, \country{USA}}}

\abstract{Despite the mounting anticipation for the quantum revolution, the success of Quantum Machine Learning (QML) in the Noisy Intermediate-Scale Quantum (NISQ) era hinges on a largely unexplored factor: the generalization error bound, a cornerstone of robust and reliable machine learning models. Current QML research, while exploring novel algorithms and applications extensively, is predominantly situated in the context of noise-free, ideal quantum computers. However, Quantum Circuit (QC) operations in NISQ-era devices are susceptible to various noise sources and errors. In this article, we conduct a Systematic Mapping Study (SMS) to explore the state-of-the-art generalization bound for supervised QML in NISQ-era and analyze the latest practices in the field. Our study systematically summarizes the existing computational platforms with quantum hardware, datasets, optimization techniques, and the common properties of the bounds found in the literature. We further present the performance accuracy of various approaches in classical benchmark datasets like the MNIST and IRIS datasets. The SMS also highlights the limitations and challenges in QML in the NISQ era and discusses future research directions to advance the field. Using a detailed Boolean operators query in five reliable indexers, we collected $544$ papers and filtered them to a small set of $37$ relevant articles. This filtration was done following the best practice of SMS with well-defined research questions and inclusion and exclusion criteria.}




\maketitle
\nomenclature[01]{$\mathcal{D}$}{A dataset $\mathcal{D} =$ \{$(\ x_1,y_1), \dots ,(\ x_N,y_N)$\}; vector of elements $(\ x_i,y_i)\ $}
\nomenclature[03]{$\mathcal{X}$}{ Input spaces whose elements are input vectors $x \in \mathcal{X}$}
\nomenclature[04]{$\mathcal{Y}$}{Output space whose elements are target values $y \in \mathcal{Y}$}
\nomenclature[05]{$f$}{An unknown target function to learn that maps $\mathcal{X}$ to $\mathcal{Y}, f:\mathcal{X} \rightarrow \mathcal{Y}$}
\nomenclature[06]{\(x \)}{The input vector \(x \in \mathcal{X},\text{ and }x_i\in \mathbb{R}^d\)}
\nomenclature[07]{\(y\)}{The target vector \(y \in \mathcal{Y}, \text{ and }y_i\in \mathbb{R}\)}
\nomenclature{\(\{\cdot\}\)}{Set}
\nomenclature[08]{\(\mathcal{A}\)}{ Learning algorithm}
\nomenclature[10]{\(g\)}{Final hypothesis selected by a learning algorithm \(\mathcal{A}\)}
\nomenclature[09]{\(\mathcal{H}\)}{Hypothesis set}
\nomenclature[11]{\(h\)}{A hypothesis \(h\in \mathcal{H}\)}
\nomenclature[02]{\(N\)}{Total number of samples in dataset \(\mathcal{D}\)}
\nomenclature[12]{\(\mathbb{E}\)}{Out of sample error or test error for hypothesis \(g\)}
\nomenclature[13]{\(\mathbb{\hat E}\)}{In sample error or training error for hypothesis \(g\)}
\nomenclature{\(\mathbb{P}[\cdot]\)}{Event probability}
\nomenclature{\(\llbracket \cdot \rrbracket\)}{Indicator function that evaluates to 1 if the argument is true, else evaluates to 0}
\nomenclature[14]{\(x_i\)}{\(i\)th training sample, \(x_i \in \mathcal{X}\)}
\nomenclature[15]{\(y_i\)}{\(i\)th training label, \(y_i \in \mathcal{Y}\)}
\nomenclature[16]{\(\hat y_i\)}{Estimate of ith label \(y_i\)}
\nomenclature{|\(\cdot\)|}{ Absolute value}
\nomenclature{\(\varepsilon\)}{Margin of error tolerance in approximating a unknown target function\(f\)}
\nomenclature{$N_y$}{ Number of classes in target set}
\nomenclature{$t_n$}{$n$-th quantum gate}
\nomenclature[17]{\(\theta\)}{Trainable vector of parameters}
\nomenclature{\(\ket{\cdot}\)}{A quantum state. Often a representation for a vector in Hilbert space }
\nomenclature{\(\bra{\cdot}\)}{The bra vector in Hilbert space that is the complex conjugate of ket vector \(\ket{.}\)}
\nomenclature{\(\mathcal{M}\)}{A measurement operator}
\nomenclature{\(\langle \cdot \rangle\)}{Measurement of an operator}
\nomenclature{\(U(x,\theta)\)}{A quantum circuit parameterized by input vector \(x \in \mathcal{X}\) and trainable parameters \(\theta \in \mathbb{R}^d\)}
\nomenclature{\(\ket{0}^{\otimes n}\)}{A quantum circuit with \(n\) qubits initialized at \(\ket{0}\) state}
\nomenclature{\(d_{vc}\)}{VC Dimension of hypothesis set}
\nomenclature{\( \delta \)}{Confidence Parameter}
\nomenclature{\(n\)}{Number of qubits}
\nomenclature{\(C_L\)}{Quantum Circuit depth}
\nomenclature[14]{\(d\)}{Feature space dimension}
\nomenclature{\(\Lambda\)}{Regularization Parameter, $\Lambda > 0$}
\nomenclature{\(\eta\)}{Margin parameter}
\nomenclature{\(\mathbb{E}\llbracket \cdot \rrbracket \)}{Expectation values}
\nomenclature{\(\rho\)}{Density matrix of \(\ket{\psi}\)}
\nomenclature{\(Tr\llbracket \cdot \rrbracket \)}{Kernel trace}
\nomenclature{\(\hat{p}\)}{Empirical distribution}
\nomenclature{\(\zeta\)}{Upper bound on Frobenius norm}
\nomenclature{\(T\)}{Number of trainable quantum circuit gates}
\nomenclature{\(p\)}{Probability of an event.}
\nomenclature[15]{\(w\)}{Learnable parameters}
\nomenclature{\(\xi\)}{Slack variable for SVM}
\nomenclature{\(\hat{W}\)}{Noisy quantum kernel matrix}
\nomenclature{\(P(x)\)}{Unknown prior probability}
\nomenclature{$L$}{The Lipschitz constant of the loss function} 
\nomenclature{$M$}{ A machine learning model} 
\nomenclature{$G$}{ Total number of gates for classical data encoding}
\nomenclature{$\Theta$}{Parameter space of a model} 
\nomenclature{$c_d$}{dimensional constant}
\nomenclature{$M_2$}{Hölder constant} 
\nomenclature{$\alpha$}{Smoothness property of Loss function in its first argument satisfying $M_2$}
\nomenclature{$l$}{ A non-negative number for model selection}
\nomenclature{$d_{\gamma, N}^{(l)}$}{Effective dimension of the discretized model $M_{\Theta}^{(l)}$} 
\nomenclature{$B$}{A positive integer that defines the loss function range}
\nomenclature{$R(\cdot)$}{ Risk of the learned function}
\nomenclature{c}{Non negative scalar}
\nomenclature{$\|g\|^2$}{Norm of the function $g$ in the RKHS}
\nomenclature{$\mathbb{E}_g$}{Generalization error}
\nomenclature{$\sigma^2$}{Variance of label noise}
\nomenclature[22]{$\Psi$}{Output state of Quantum Neural Network (QNN)}
\nomenclature[23]{$\Phi$}{Input state to QNN}
\nomenclature{$\mathbbm{1}$}{Indicator Function}
\nomenclature{$K$}{number of parameterized local completely positive and trace-preserving maps}
\nomenclature{$S$}{Number of structures allowed for QNN}
\nomenclature{$L_c$}{Training steps complexity for QML model to generalize}
\nomenclature{$R_1$}{Minimal degree of freedom rank for isometric unitary. }
\nomenclature{$R_\infty$}{Maximum degree of freedom rank for isometric unitary. }
\nomenclature{$C(\mathcal{H})$}{Complexity measure of $\mathcal{H}$}
% \printnomenclature \label{sec:nomen}

\section{Introduction}\label{sec:intro}
In the field of machine learning, the fundamental theory of learning from data~\citep{yasirlfdata}, it is defined that any model aims to learn an unknown target function $f:\mathcal{X} \rightarrow \mathcal{Y}$ from a dataset $\mathcal{D} = \{(x_1,y_1),\ldots,(x_N,y_N)\}$ using a learning algorithm $\mathcal{A}$, where $\mathcal{X}$ is an input space with $d$ feature space dimensions, $\mathcal{Y}$ is an output space, \(x_i \in \mathcal{X},\text{ and }x_i\in \mathbb{R}^d\), \(y_i \in \mathcal{Y} \text{ and }y_i\in \mathbb{R}\), \( i = 1, \dots, N, \) and $N$ is the total number of samples in the dataset $\mathcal{D}$. The learning algorithm $\mathcal{A}$ selects a hypothesis $g$ from a hypothesis set $\mathcal{H}$ that best approximates the unknown target function $f$.
In classical machine learning, the generalization error or generalization gap is the difference between a model's performance on the training data and unseen data~\citep{nadeau1999inference,jakubovitz2019generalization,emami2020generalization}. The true error $\mathbb{E}(g) = \mathbb{P}[g(x) \neq f(x)]$ is the expected difference between the hypothesis output $g(x)$ and the actual output $f(x) = y$. In this context, $\mathbb{P}[\cdot]$ is the probability of an event. Similarly, the empirical error $\mathbb{\hat{E}}(g) = \frac{1}{N}\sum_{i=1}^N\llbracket g(x_i) \neq f(x_i) \rrbracket$ is the average difference between true label $y_i = f(x_i)$ and predicted label $\hat{y_i}  = g(x_i)$ over the dataset. Here $g(x)$ is the model's current best approximation of $f$.
A bound on the generalization error provides an upper limit to the model's error rate on unseen data, assuming that the unseen data is drawn from the same distribution as the training data. Generalization bound is typically derived using statistical learning theory and depends on factors such as the complexity of the model (hypothesis class), the number of examples in the training data, and the randomness in the data generation process. Hoeffding's inequality~\citep{hoeffding1994probability} gives a common form of the generalization bound in classical machine learning. Hoeffding's inequality provides an error bound for $f$ based on $\mathcal{D}$ \citep{hoeffdingcs229} by giving the deviation probability of $\mathbb{\hat{E}}(g)$ from $\mathbb{E}(g)$ as function of a positive tolerance $\varepsilon$ and $N$, and is given by:
\begin{equation}\label{eq:hoeffding}
\mathbb{P} [ |  \mathbb{E}(g) - \mathbb{\hat{E}}(g)|\geq \varepsilon ] \leq 2e^{-2N\varepsilon^2}
\end{equation}

This equation states that the probability that the absolute difference between the in-sample, training, and out-of-sample, testing, errors being greater than $\varepsilon$ is less than or equal to $2e^{-2N\varepsilon^2}$. However, the effectiveness of this bound may be reduced or invalidated when the underlying random variable is affected by the noise, as is the case with NISQ devices \citep{du2021learnability,de2023limitations,hakkaku2022quantifying}. While some noise is beneficial for classical machine learning  \citep{neelakantan2015adding}, excessive noise in quantum systems can limit the performance of QML algorithms \citep{bharti2021noisy,wang2021towards}. To overcome this challenge, NISQ-era hardware-related noise and QC operations-induced noise must be considered when developing reliable  QML models.

QML is an emerging field with great promise for revolutionizing learning from data~\citep{biamonte2017quantum,arunachalam2017guest,carleo2019machine,wittek2014quantum}. It focuses on improving machine learning using quantum systems through a mathematical framework \citep{arunachalam2017guest}. Several learning models~\citep{arunachalam2017guest, martin2022quantum} have been proposed in QML, including Probably Approximately Correct (PAC) that explores how quantum resources, such as superposition and entanglement, can improve the sample complexity or computational efficiency of learning classical concepts \citep{rocchetto2019experimental}. However, our ability to harness QML capabilities is particularly impacted by the practical limitations of NISQ devices, which are presently the most advanced quantum computers available~\citep {preskill2018quantum,bharti2021noisy}.
Variational quantum computing~\citep{cerezo2021variational} uses imperfect NISQ-era devices for computation. The variational quantum Circuit (VQC) model is a QML model that can be described as a quantum circuit model \citep{wittek2014quantum,schuld2021machine} and is defined as: 
\begin{equation}\label{eq:qmlmodeldef}
f_{\theta}(x) = \bra{\psi(x,\theta)} \mathcal{M} \ket{\psi(x,\theta)}
\end{equation}

with \(\ket{\psi(x,\theta)}\) being a quantum state prepared by a Parameterized Quantum Circuit (PQC) \(U(x,\theta)\) with trainable parameters $\theta$~\citep{benedetti2019parameterized}, and \(\mathcal{M}\) a measurement operator. On a system with $n$ number of qubits, the circuit \(U(x,\theta)\) initiates with a sequence of quantum gates from an initial state predominantly \(\ket{0}^{\otimes n}\)\citep{biamonte2017quantum}. The relationship between the PQC and the unitary is crucial and can be defined as follows:
\begin{equation} \label{eq:PQCunitaryrelation}
\ket{\psi(x,\theta)} = U(x,\theta)\ket{0}^{\otimes n}
\end{equation}

Though theoretically promising, studies have shown that deep quantum circuits are especially vulnerable to noise, accumulating gate errors and experiencing significant decoherence \citep{alam2022qnet,wang2022quantumnat}. In addition, the benefits of quantum kernels~\citep{wang2021towards,heyraud2022noisy,kubler2021inductive,schuld2021supervised,thanasilp2024exponential} are reduced in the presence of large system noise and a higher number of measurements \citep{huang_2021,wang2021towards,preskill2018quantum,thanasilp2024exponential}. Furthermore, the limited data available for training quantum models becomes even more challenging to work with due to the noise inherent in NISQ devices, leading to potential misdirection in learning and increasing generalization error \citep{schuld2019quantum}.

The implication is clear: algorithms implemented on NISQ devices are susceptible to considerable noise and may not work as expected \citep{wang2021towards,bharti2021noisy}, causing a divergence between theoretical predictions and empirical results. To fully exploit the potential of QML models in the NISQ era, it is crucial to develop a deep understanding of these constraints and establish a robust error bound to account for the impact of the existing system noise and hardware limitations. Consequently, a dedicated study to define and understand the Generalization Error Bound (GEB) for QML in the NISQ era is not merely a theoretical interest but a critical necessity for the practical realization of the potential of quantum computation. 

In this paper, we investigate the GEB in supervised QML and its validity in the NISQ era. Additionally, we seek to explore the types of algorithms used in QML research, the platform of choice for these algorithm implementations, optimization techniques, datasets, and whether most of the work is theoretical or experimental. Further, we investigate the effectiveness of these QML models by presenting their performance metrics on classical benchmark datasets such as MNIST, Fashion MNIST, and IRIS, along with the number of classes analyzed and the context of the experiments, whether conducted in noisy or ideal settings, as described in~\ref{subsubsec:performace}. We conducted a Systematic Literature Review (SLR) designed as an SMS to achieve these objectives. Research has demonstrated that SMS is a valuable tool for organizing and categorizing existing discoveries while also identifying limitations and gaps for improvements~\citep{kitchenham2011using}. This paper is structured as follows: Section ~\ref{sec:method} provides an overview of the methodology for SLR, including its protocols. Section~\ref{sec:result} presents the result of this SLR, and section~\ref{sec:discussion} provides the discussion. Finally, in section~\ref{sec:conclusion}, we conclude this review with a summary of our findings and their implications.

\section{Methodology}\label{sec:method}
SLR in this study followed the ~\cite{kitchenham2011using} process, which consists of three phases: planning, conducting, and reporting. This approach helped us to systematically identify, analyze, evaluate, and interpret the literature to answer the proposed research questions. In the planning phase, we selected the database for the literature search and defined research questions along with the inclusion and exclusion criteria. We also designed the boolean algebra for the search query. The conducting phase involved identifying relevant articles based on the defined protocols. Finally, the results were interpreted and reported in a structured format in the reporting stage.

For this review, we used two applications: \say{Publish or Perish}\footnote{https://harzing.com/resources/publish-or-perish/} and \say{Zotero} \footnote{https://www.zotero.org/}. \say{Publish or Perish} was used for searching the literature across multiple platforms, while \say{Zotero} was used for organizing the literature, checking for duplicates, and generating a bibliography. Additionally, we used Microsoft Excel to keep track of our progress during various stages of the review.

\begin{table}
\caption{Search results from various sources}
\label{tab:search}
\centering
\begin{tabularx}{0.5\textwidth}{|l|X|l|}
\hline
\textbf{Source}  & \textbf{Field} & \textbf{Result Count} \\
\hline
Google Scholar  & All Fields & 183 \\
\hline
ACM Digital Library &  Title, Abstract & 108 \\
\hline
Semantic Scholar & All Fields & 106 \\
\hline
Scopus &  Title, Abstract, Keywords & 79 \\
\hline
IEEE Xplore& Abstract & 58 \\

\hline
\multicolumn{2}{|l|}{Snow Balling}  & 10 \\
\hline
\multicolumn{2}{|l|}{Duplicates}  & 118 \\
\hline
\multicolumn{2}{|l|}{Article published before 2010}  & 26 \\
\hline
\multicolumn{2}{|l|}{Total}  & 688/544 \\
\hline
\end{tabularx}
\end{table}
\begin{figure}[t]
    \centering
    \includegraphics[width = 0.45\textwidth,trim=0 0 0 0,clip]{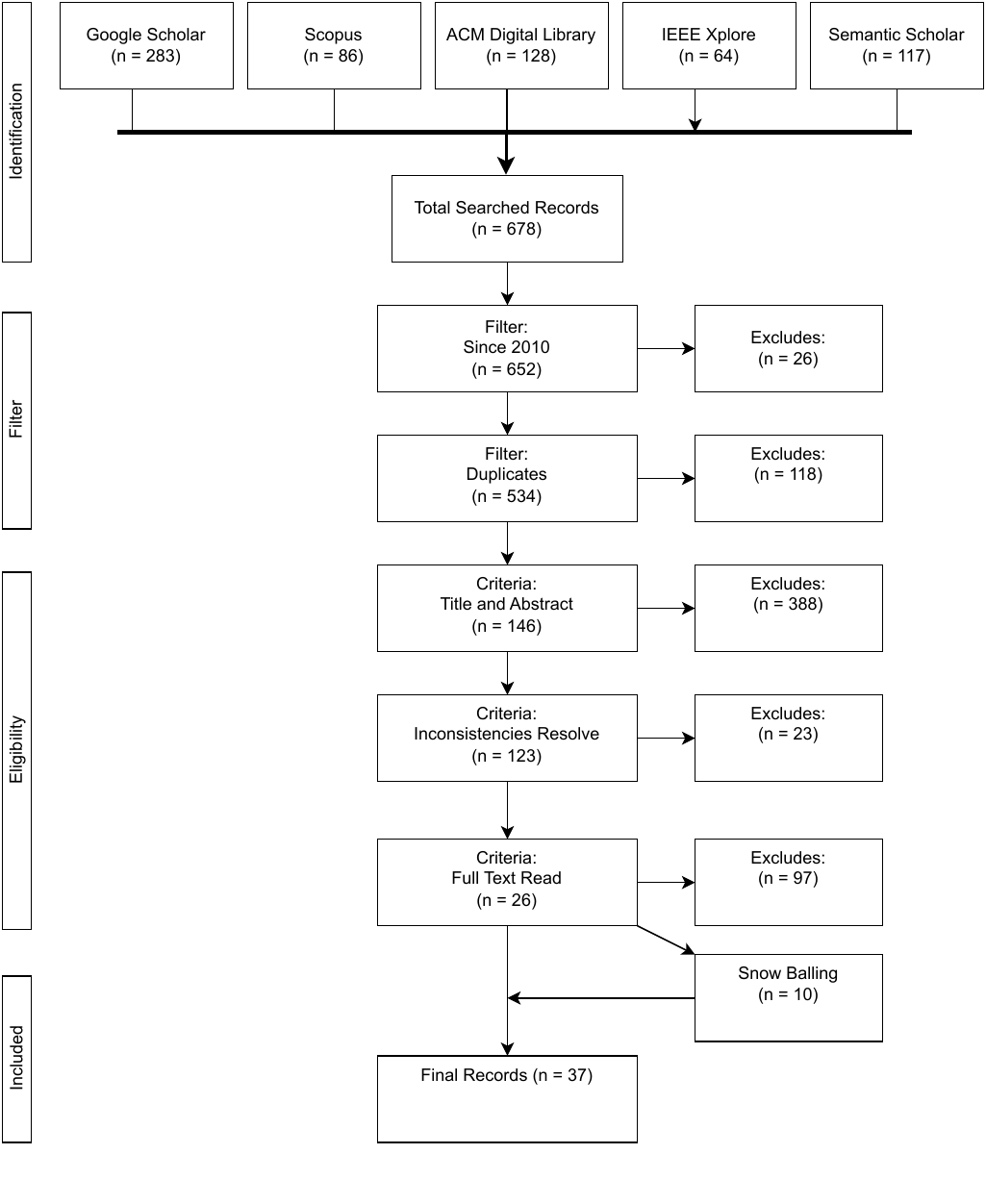}
    \caption{The Prisma diagram provides the literature counts at various stages of the SLR process. From the collected paper count of $688$, the filtration steps excluded $144$ papers, and the eligibility steps excluded $507$ papers, resulting in $37$ papers for the analysis.}
    \label{fig:prisma}
\end{figure}

We conducted a systematic literature search across five academic databases to identify relevant papers for our study. The databases included Google Scholar, Scopus, ACM Digital Library, IEEE Xplore, and Semantic Scholar. We collected a total of $678$ articles from various platforms, which, after accounting for overlaps and duplicates, resulted in $534$ unique publications. We added $10$ articles that did not appear in the initial articles but were relevant to this study. These articles were hand-picked during the snowballing process. Among databases, Google Scholar had the highest number of relevant papers, totaling $183$, after removing duplicates, representing $33.7\%$ of the entire collection. ACM Digital Library and Semantic Scholar followed, with $108 (20\%)$ and $106 (19.5\%)$ papers, respectively. Scopus sourced $79$ articles, making up $14.54\%$, while IEEE Xplore contributed $58$ articles, which was $10.68\%$ of the total. About $1.6\%$ of articles were manually added. Additionally, we used query strings \say{ `Quantum' OR `quantum' AND `Quantum Machine Learning' AND `error bound' AND `noisy' AND `NISQ' } or \say{ `Quantum Machine Learning' AND `error bound' } to search for the relevant literature from $2010-2023$. At the end of the study, we identified $37$ articles that met the inclusion criteria. We summarized the identification and selection process using the PRISMA diagram in Figure~\ref{fig:prisma}. 
\begin{figure*}[ht]
     \begin{subfigure}[b]{0.5\textwidth}
         \centering
        \includegraphics[width=\textwidth]{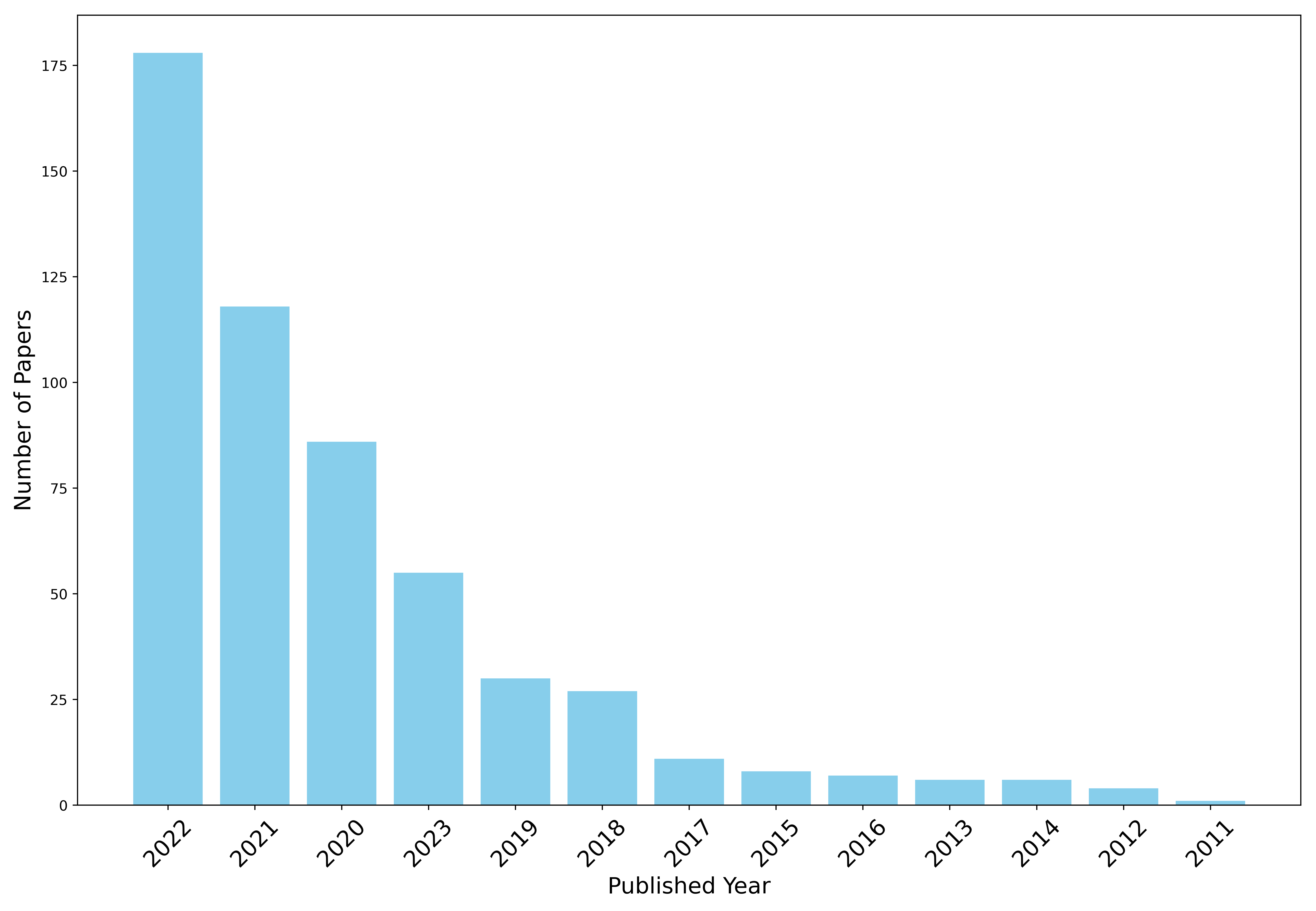}
        \caption{}
        \label{fig:search_results}
    \end{subfigure}
        \hfill
    \begin{subfigure}[b]{0.5\textwidth}
             \centering
             \includegraphics[width=\textwidth]{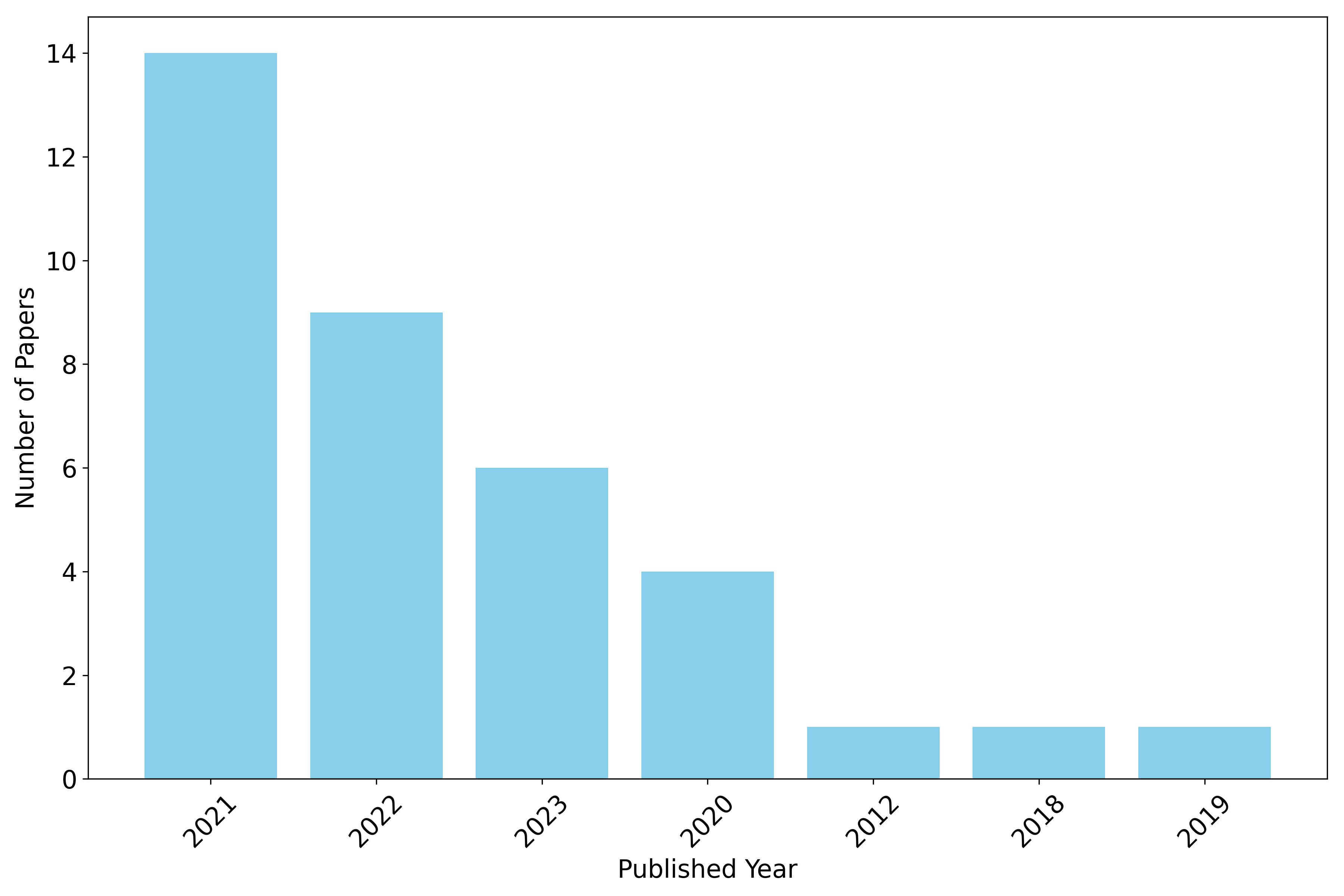}
             \caption{}
             \label{fig:paperperyear}
    \end{subfigure}
        \caption{ a) Bar graph presenting the number of publications per year from 2010 to 2023 among initially extracted papers. It is evident that the number of publications has increased over the years, mainly in the last 5 years. b) Bar graph for the number of papers extracted per year among the final papers.}
\label{fig:PublicationbyYear}
\end{figure*}

\begin{figure*} [t]   
    \begin{subfigure}[b]{0.5\textwidth}
             \centering
             \includegraphics[width=\textwidth]{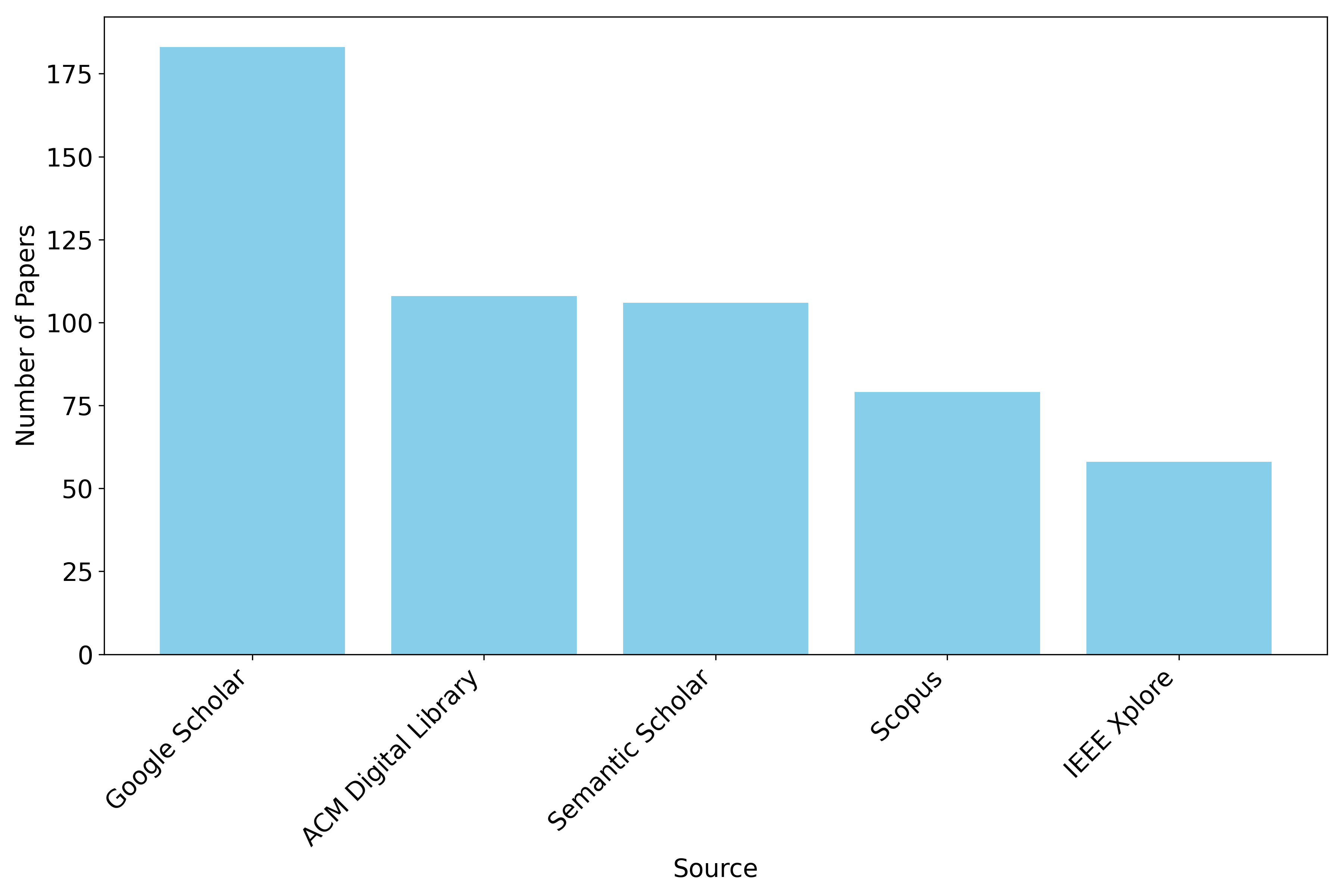}
            \caption{}
            \label{fig:authorsPerPaper}  
    \end{subfigure}
    \hfill
    \begin{subfigure}[b]{0.5\textwidth}
             \centering
             \includegraphics[width=\textwidth]{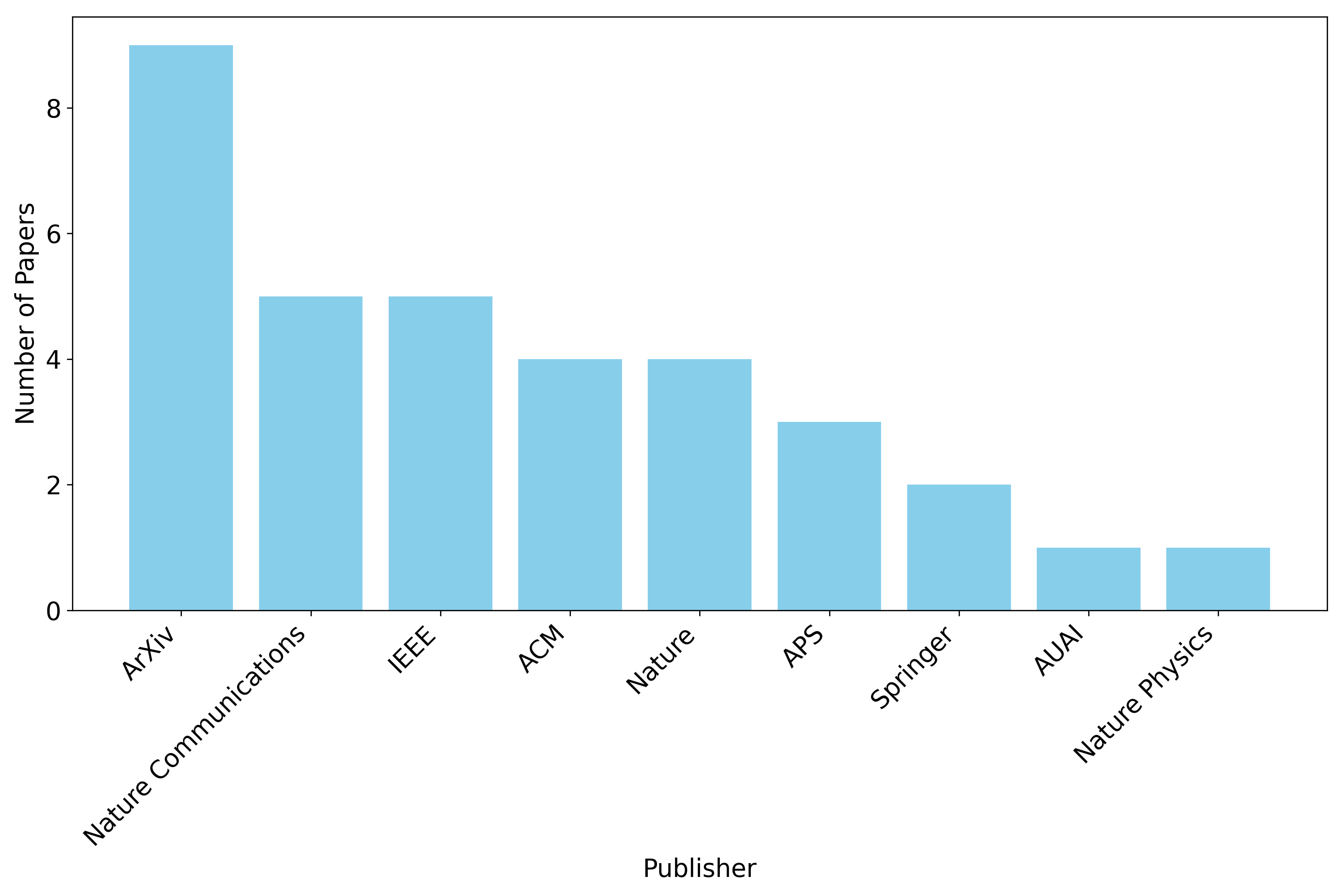}
             \caption{}
             \label{fig:Paperpersource}
    \end{subfigure}

\caption{a) Figure showing the number of papers extracted from each database source. Google Scholar was the most significant source for this study. b) Bar graph for the number of papers extracted from each publisher among the final papers. We can see that the majority of the papers were published in ArXiv. This graph represents the final papers after the filtration process. Naturally, it is expected that the platform with free access to full-text articles would have the most papers.}
\label{fig:paper_per_source}
\end{figure*}

\begin{figure*}[t]
    \begin{subfigure}[b]{0.5\textwidth}
         \centering
        \includegraphics[width=\textwidth]{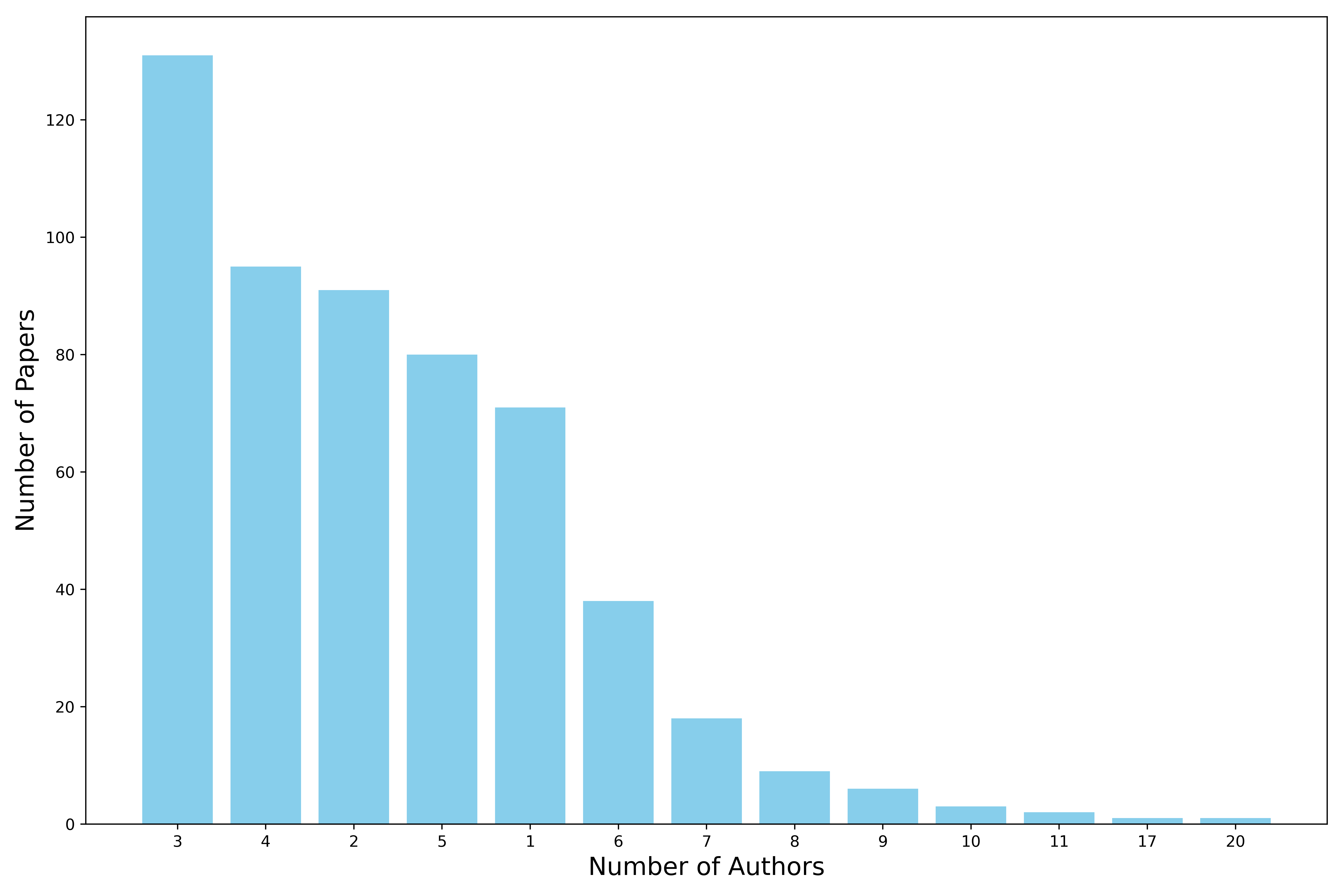}
        \caption{}
        \label{fig:paperperpublish}
    \end{subfigure}
        \hfill
    \begin{subfigure}[b]{0.5\textwidth}
             \centering
             \includegraphics[width=\textwidth]{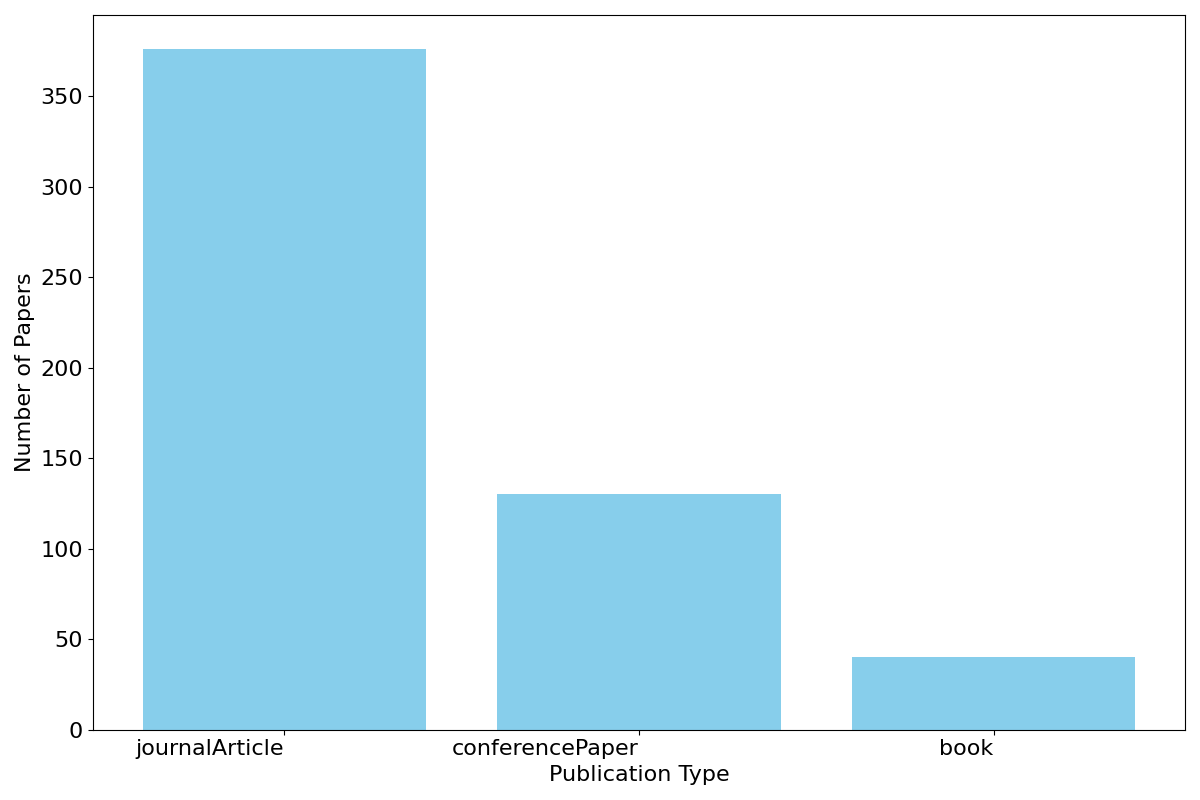}
             \caption{}
             \label{fig:paper_per_type}
    \end{subfigure}
 \caption{a) Bar graph for a count of authors per paper. For most papers, we can see a collective research effort trend of $2-5$ authors per paper. b) Bar graph for the number of papers per publication type. The majority of the papers were journal articles. This suggests a preference for formal, peer-reviewed research channels over conference papers.}
 \label{fig:PapersPerType}
\end{figure*}

Figures~\ref{fig:PublicationbyYear},\ref{fig:paper_per_source} and \ref{fig:PapersPerType} reveal key trends in the field’s research practices. Fig.~\ref{fig:PublicationbyYear} shows an increase in publications number as the year progresses, suggesting a growing research interest in the field. However, our rigorous selection process led to the inclusion of only a few of these articles. It's possible that although there has been growing interest in QML recently, there are very few works that specifically focus on studying the generalization bound. Fig.~\ref{fig:paper_per_source} ’s analysis of the databases shows Google Scholar as the most significant source. Lastly, Fig.~\ref{fig:PapersPerType} indicates a common trend of $2-5$ authors per paper, pointing to collaboration in the field, and highlights journal articles as the favored publication type, suggesting a preference for formal, peer-reviewed research channels over conference papers. This analysis highlights the importance of research quantity and quality, the majority of collaborative efforts, and a tendency toward standard publication types.

\subsection{Planning}\label{subsec:planning}
The study primarily focuses on current practice in QML, possible developments, and the difficulties that accompany it, especially regarding the generalization error bound. We formed a team of six members to ensure the study maintains rigor and avoids biases. These members had specialized backgrounds in machine learning, quantum computing, and quantum machine learning. This collective expertise ensured comprehensive coverage and allowed us to view the articles from multiple directions.

The QML field has gathered much attention in the last decade or so. For this reason, we limited our search from $2010$ to $2023$\footnote{Some of the recently published articles were included as part of the snowballing phase}. We selected the dataset mentioned above because these databases provided access to a large number of articles with full-length searches or customized searches of an article and were available for free via the licenses held by the University. Once the relevant research papers were collected, the next step involved meticulously extracting the information. We focused on details like the authors, publication year, venue, research approach, tools and techniques employed, a platform used for quantum computation, outcomes, challenges, datasets and data encoding strategy, optimization technique, and further research suggestions. This structured extraction method ensured that the data we gathered was comprehensive and easy to interpret and analyze.

\subsection{Research Questions}\label{subsec:RQ}
In this article, we focused on answering the following questions:
\begin{enumerate}
    \item RQ1: What is the current state-of-the-art Generalization Error Bound for Quantum Machine Learning applied to Noisy Intermediate-Scale Quantum (NISQ) devices?
    \item RQ2: What are the current standard practices in QML in the context of NISQ?
    \item RQ2a. Is the majority of the research focused on theoretical, empirical, or other approaches?
    \item RQ2b. How is success measured in QML research: complexity, accuracy, training time, or other metrics?
    \item RQ2c. What types of datasets are commonly used in QML research: real, synthetic, or others?
    \item RQ3.  What computing platforms/devices are used in the experiments?
\end{enumerate}

\subsection{Inclusion and Exclusion Criteria}\label{subsec:IncExc}
We defined the following inclusion and exclusion criteria for filtering the extracted article. The filtration process was divided into three phases: Phase 1, Phase 2, and Phase 3 (which shall be discussed later). Any article satisfying at least two inclusion criteria in Phase 1 moved to Phase 2. Also, any article satisfying at least one exclusion criterion was directly excluded in Phase 1.

Inclusion Criteria:
\begin{enumerate}
    \item Topic: Papers focused on QML in the context of the NISQ era.
    \item Error Bounds: Papers discussing Hoeffding's error bound or related error bounds or noise for QML in NISQ.
    \item Research Approach: Papers presenting theoretical, empirical, or other research approaches relevant to the research questions.
\end{enumerate}

Exclusion Criteria:
\begin{enumerate}
    \item Language: Papers not published in English.
    \item Topic: Papers not focused on QML or NISQ devices.
    \item Error Bounds: Papers that do not discuss generalization error bound, noise in NISQ, or related error bounds.
    \item Relevance: Papers that do not provide sufficient information or context to address the research questions.
    \item Publication Type: Non-peer-reviewed articles, such as opinion pieces, editorials, or preprints without substantial evidence or contribution to the field or Book.
    \item Duplicates: Papers that have already been included in the review.	
\end{enumerate}

\begin{table*}[htbp]
\caption{List of final papers }
\label{tab:finalPapers}
\begin{tabular}{|>{\raggedright\arraybackslash}p{5cm}|>{\raggedright\arraybackslash}p{3.3cm}|>{\raggedright\arraybackslash}p{1.5cm}|>{\raggedright\arraybackslash}p{2cm}|>{\raggedright\arraybackslash}p{2cm}|}
\toprule
Title & Reference & Published Year & Source & Publisher \\
\hline
Robust Classification with Adiabatic Quantum Optimization & \cite{denchev2012robust} & 2012 & Semantic Scholar & ArXiv \\ \hline
Implementable Quantum Classifier for Nonlinear Data & \cite{du2018implementable} & 2018 & Google Scholar & ArXiv \\ \hline
Towards quantum machine learning with tensor networks & \cite{huggins2019towards} & 2019 & Google Scholar & ArXiv \\ \hline
Error-mitigated data-driven circuit learning on noisy quantum hardware & \cite{hamilton2020error} & 2020 & Scopus & Springer \\ \hline
Quantum classifier with tailored quantum kernel & \cite{blank2020quantum} & 2020 & Google Scholar & Nature \\ \hline
Quantum Error Mitigation With Artificial Neural Network & \cite{kim2020quantum} & 2020 & IEEE & IEEE \\ \hline
The Born supremacy: quantum advantage and training of an Ising Born machine & \cite{coyle2020born} & 2020 & Scopus & NPJ Quantum Inform. \\ \hline
Layerwise learning for quantum neural networks & \cite{skolik2021layerwise} & 2021 & Scopus & Springer \\ \hline
On the expressibility and overfitting of quantum circuit learning & \cite{chen2021expressibility} & 2021 & ACM & Quantum (ACM) \\ \hline
A rigorous and robust quantum speed-up in supervised machine learning & \cite{liu2021rigorous} & 2021 & Scopus & Nature Physics \\ \hline
Robust quantum classifier with minimal overhead & \cite{park2021robust} & 2021 & Scopus & IEEE \\ \hline
I-QER: An Intelligent Approach Towards Quantum Error Reduction & \cite{basu2022qer} & 2021 & ACM & ACM \\ \hline
Can Noise on Qubits Be Learned in Quantum Neural Network? A Case Study on QuantumFlow & \cite{liang2021can} & 2021 & Scopus & IEEE \\ \hline
Power of data in quantum machine learning & \cite{huang_2021} & 2021 & Scopus & Nature communications \\ \hline
Generalization in Quantum Machine Learning: a Quantum Information Perspective & \cite{banchi2021generalization} & 2021 & Semantic Scholar & APS \\
\bottomrule
\end{tabular}
\end{table*}

\begin{table*}[htbp]
\begin{tabular}{|>{\raggedright\arraybackslash}p{5cm}|>{\raggedright\arraybackslash}p{3cm}|>{\raggedright\arraybackslash}p{1.5cm}|>{\raggedright\arraybackslash}p{2cm}|>{\raggedright\arraybackslash}p{2cm}|}
% \toprule
\multicolumn{5}{c}{\texttt{Continued from previous table, Table~\ref{tab:finalPapers}}} \\
\hline
Quantum One-class Classification With a Distance-based Classifier & \cite{de2021quantum} & 2021 & IEEE & IEEE \\ \hline
Noise-induced barren plateaus in variational quantum algorithms & \cite{wang2021noise} & 2021 & Google Scholar & Nature \\ \hline
Encoding-dependent generalization bounds for parametrized quantum circuits. & \cite{caro2021encoding} & 2021 & Snow Balling & Quantum \\ \hline
The power of quantum neural networks & \cite{abbas2021power} & 2021 & Snow Balling & Nature Communication \\ \hline
The inductive bias of quantum kernels & \cite{kubler2021inductive} & 2021 & Snow Balling & NeurIPS \\ \hline
Towards understanding the power of quantum kernels in the NISQ era & \cite{wang2021towards} & 2021 & Scopus & Quantum \\ \hline
Problem-Dependent Power of Quantum Neural Networks on Multi-Class Classification & \cite{du2022problem} & 2022 & Snow Balling & ArXiv \\ \hline
Theoretical error performance analysis for variational quantum circuit based functional regression & \cite{qi2023theoretical} & 2022 & Google Scholar & Nature \\ \hline
QOC: quantum on-chip training with parameter shift and gradient pruning & \cite{wang2022qoc} & 2022 & Semantic Scholar & ACM \\ \hline
Quantum Perceptron Revisited: Computational-Statistical Tradeoffs & \cite{roget2022quantum} & 2022 & Scopus & AUAI \\ \hline
Implementation and Empirical Evaluation of a Quantum Machine Learning Pipeline for Local Classification & \cite{zardini2022implementation} & 2022 & Google Scholar & ArXiv \\ \hline
Noisy quantum kernel machines & \cite{heyraud2022noisy} & 2022 & Google Scholar & APS \\ \hline
A kernel-based quantum random forest for improved classification & \cite{srikumar2022kernel} & 2022 & Google Scholar & ArXiv \\ \hline
Bandwidth Enables Generalization in Quantum Kernel Models & \cite{canatar2022bandwidth} & 2022 & Snow Balling & ArXiv \\ \hline
The Dilemma of Quantum Neural Networks & \cite{qian2022dilemma} & 2022 & IEEE & IEEE \\
\bottomrule
\end{tabular}
\end{table*}

\begin{table*}[htbp]
\begin{tabular}{|>{\raggedright\arraybackslash}p{5cm}|>{\raggedright\arraybackslash}p{3cm}|>{\raggedright\arraybackslash}p{1.5cm}|>{\raggedright\arraybackslash}p{2cm}|>{\raggedright\arraybackslash}p{2cm}|}
% \toprule
\multicolumn{5}{c}{\texttt{Continued from previous table, Table~\ref{tab:finalPapers}}} \\ 
\hline
Generalization with quantum geometry for learning unitaries & \cite{haug2023generalization} & 2023 & Snow Balling & ArXiv \\ \hline
Ensemble-learning variational shallow-circuit quantum classifiers & \cite{li2023ensemble} & 2023 & Google Scholar & ArXiv \\ \hline
Generalization in quantum machine learning from few training data & \cite{caro2022generalization} & 2023 & Google Scholar & Nature \\ \hline
Understanding quantum machine learning also requires rethinking generalization & \cite{gil2023understanding} & 2023 & Snow Balling & ArXiv \\ \hline
Quantum machine learning beyond kernel methods & \cite{jerbi2023quantum} & 2023 & Snow Balling & Nature Communications \\ \hline
Out-of-distribution generalization for learning quantum dynamics & \cite{caro2023out} & 2023 & Snow Balling & Nature Communications \\ \hline
Dynamical simulation via quantum machine learning with provable generalization & \cite{gibbs2024dynamical} & 2024 & Snow Balling & APS \\
\bottomrule
\end{tabular}
% \end{adjustbox}
\end{table*}
% \end{longtable}
\subsection{Search Process and Query String}\label{subsec:QueryString}
The initial phase of this review involved gathering relevant publications as described previously. The search process consisted of formulating search queries tailored to each database and successively filtering out papers based on inclusion and exclusion criteria until final selections were made. We defined the following keywords to aid the search process: \textit{Quantum Machine Learning, Generalization Error; NISQ Devices, Quantum Circuits.} With these keywords defined, we constructed search queries for various platforms as follows:
\begin{enumerate}
    \item Google Scholar: (\textit{All Fields}: Quantum OR quantum AND Quantum Machine Learning AND error bound AND data AND noisy AND NISQ. Date: $2010-2023$)
    \item Scopus: (\textit{Title, Abstract and Keywords}: Quantum Machine Learning AND noisy AND NISQ AND PUBYEAR  $> 2010$ AND PUBYEAR $<= 2023$)
    \item ACM Digital Library: (\textit{Title}: quantum) AND (\textit{Abstract}: quantum machine learning) AND (\textit{E-Publication Date}: ($01/01/2010$ TO $03/31/2023$))
    \item IEEE Xplore: (\textit{All Metadata}: quantum machine learning) AND (\textit{Abstract}: quantum machine learning) AND (\textit{Abstract}: bound) . Date:$2010-2023$.
    \item Semantic Scholar:  (\textit{All Fields}: Quantum Machine Learning AND bound AND NISQ, Date: $2010-2023$)
\end{enumerate}

From these searches, we identified $688$ papers. Upon removing duplicates, we had $534$ unique articles. Table~\ref{tab:search} details the search counts for each platform based on the formulated queries and fields. Figure~[\ref{fig:PublicationbyYear},\ref{fig:paper_per_source},\ref{fig:PapersPerType}] visually illustrates the publication rates, source-specific paper counts, and distribution of paper types.

The filtration and information extraction were divided into the following phases:\\
\textbf{Phase 1:} At this stage, each paper was evaluated by two members based solely on its title, abstract, and keywords following the inclusion/exclusion criteria. Papers receiving at least one approval proceeded to phase $2$. During this process, we removed $388$ irrelevant papers and proceeded with $146$ for phase $2$.\\
\textbf{Phase 2:} In this phase, we addressed any inconsistencies that had occurred in phase 1. We flagged that the decision for an article is inconsistent if both the reviewers had different decisions for a paper during the phase 1 review. Of the 146 papers, 46 had inconsistent decisions. Different team members who hadn't previously worked on inconsistency papers re-evaluated each of these papers using the inclusion/exclusion criteria. This phase concluded with $23$ of the $46$ disputed papers advancing to phase $3$, totaling $123$ papers.\\
\textbf{Phase 3}: Each team member was assigned a subset of papers and was tasked with thoroughly reading assigned papers and determining their suitability for the review. They were thereupon asked to decide if the paper should be included in the study. By the end of this phase, $27$ papers were deemed pertinent to our study.\\
\textbf{Phase 4}: Those who undertook comprehensive paper assessments in Phase 3 were then responsible for extracting key data based on a predefined coding schema.\\
\textbf{Snow Balling}: This process included adding the papers that, for some reason, didn't show up in the search result or were filtered during the above phase but are relevant to the research. We added $10$ papers manually during the review process.

With the final papers selected, we extracted the relevant information from each paper. Table~\ref{tab:finalPapers} presents the final list of papers. In the next section, we present the results of our analysis of these papers.

\section{Result}\label{sec:result}
In this section, we present the results of our SLR. We begin by discussing the datasets and optimization techniques used in QML research. We then present the performance metrics of the QML models and the platforms used for the experiments. We also discuss the research approach, generalization and other relatable error bounds, and the experimental and theoretical nature of the research.

\subsection{Dataset}\label{subsub:dataset}
\begin{table*}[htbp]
    \centering
    \caption{Comparative Overview of Datasets and Optimization Techniques }
    \label{tab:datasetandOpt}
     \begin{tabular}{|p{3cm}|p{5.7cm}|p{6cm}|}
    \hline
    Reference & Type of Dataset & Optimization Technique \\
    \hline
    \cite{basu2022qer}  & Synthetic: Tabular with quantum gate as features & Grid Search \\
    \hline
    \cite{blank2020quantum}  & Synthetic: Quantum states parametrized by angles & Optimize1qGates (Qiskit) \\
    \hline
    \cite{du2018implementable}  & Synthetic: Linear and nonlinear dataset generated following~\cite{havlivcek2019supervised} & Parameter-shift rule \\
    \hline
    \cite{chen2021expressibility}  & Real: Iris dataset and synthetic: Non-linear  & Backpropagation (Simulator) following~\cite{watabe2019quantum} \\
    \hline
    \cite{srikumar2022kernel}  & Real: Fashion MNIST, Breast Cancer and Heart Diseases & Nystrom Approximation \\
    \hline
    \cite{heyraud2022noisy} & Real: MNIST & Least-square loss function minimization \\
    \hline
    \cite{zardini2022implementation}  & Real: UCI and Iris & Hamming distance optimization \\
    \hline
    \cite{qi2023theoretical}  & Real: MNIST & SGD with Adam optimizer \\
    \hline
    \cite{li2023ensemble} & Real: MNIST and synthetic: phase recognition & Automatic differentiation with Adam \\
    \hline
    \cite{caro2022generalization} & Synthetic: Ground states phase & SPSA with Matrix Product State \\
    \hline
    \cite{roget2022quantum} & Real: Iris and synthetic: following~\cite{mohri2018foundations} & Gradient descent  \\
    \hline
    \cite{huang_2021}  & Real: Fashion-MNIST and synthetic: Engineered dataset & Gradient Descent \\
    \hline
    \cite{liu2021rigorous}  & Synthetic: 2D points with hyperplane distance & Convex quadratic optimization \\
    \hline
    \cite{skolik2021layerwise}  & Real: MNIST & Parameter-shift with binary cross-entropy \\
    \hline
    \cite{liang2021can}  & Real: MNIST & Qubit Mapping \\
    \hline
    \cite{wang2021towards}  & Real: Fashion-MNIST and synthetic: Engineered dataset & Grid search (Regularization parameter) \\
    \hline
    \cite{coyle2020born}  & Synthetic: Engineered dataset (QCIBM) & SGD (Parameter-shift rule) \\
    \hline
    \cite{hamilton2020error} & Synthetic: 4-qubit circuit targets & Gradient-based (Adam, Parameter-shift) \\
    \hline
    \cite{huggins2019towards} & Real: MNIST & SPSA, Finite difference gradient \\
    \hline
    \cite{wang2022qoc} & Real: MNIST, Vowel-4 & SGD with Adam \\
    \hline
    \cite{banchi2021generalization}  & Synthetic: 2-Moon & Variational Quantum Info Bottleneck \\
    \hline
    \end{tabular}
    \end{table*}

    \begin{table*}[htbp]
    \begin{tabular}{|p{3cm}|p{5.7cm}|p{6cm}|}
    % \toprule
    \multicolumn{3}{c}{\texttt{Continued from previous table, Table~\ref{tab:datasetandOpt}}} \\
    \hline
    \cite{denchev2012robust} &Real: UCI and synthetic: Long-Servedio, Mease-Wyner  & Adiabatic quantum optimization \\
    \hline
    \cite{kim2020quantum} & Synthetic: Rand. quant. circuits & Gradient descent (RMSE) \\
    \hline
    \cite{de2021quantum} & Real: Iris & SGD \\
    \hline
    \cite{qian2022dilemma} & Real: Wine, MNIST & SGD, SQNGD \\
    \hline
    \cite{wang2021noise} & Synthetic: Randomly generated graphs following~\cite{erdds1959random} & Quantum Approximate Optimization Algorithm \\
    \hline
    \cite{abbas2021power} & Real: Iris (First two classes) & Cross-entropy loss with Adam \\ \hline 
    \cite{kubler2021inductive} & Synthetic: drawn from a uniform distribution on $[0,2\pi]^d$ & Mean square error with Kernel-target Alignment. \\ \hline
    \cite{canatar2022bandwidth} & Real: FMNIST, KMNIST, PLAsTiCC and Synthetic following~\cite{shaydulin2022importance} & Convex quadratic optimization \\ \hline
    \cite{jerbi2023quantum} & Real: FMNIST & Gradient descent with Adam  \\ \hline
    \cite{caro2023out} & Synthetic: Random product states & Gradient Free Nelder-Mead \\ \hline
    \cite{gibbs2024dynamical} & Synthetic: Har-random product states & Gradient descent \\ \hline
    \cite{gil2023understanding} & Synthetic: Generalized cluster Hamiltonian of $n$ qubits & Covariance Matrix Adaptation Evolution Strategy  \\ \hline %
    \cite{haug2023generalization} & Synthetic: Random product states & Gradient Descent \\ \hline
    \cite{du2022problem} & Parity and FMNIST & Gradient descent with Adagrad optimizer \\ \hline
\end{tabular}
\end{table*}

Table~\ref{tab:datasetandOpt} shows a wide diversity in both the types of datasets and optimization techniques used in QML research, offering a glimpse into the current practice of the field. The dataset types vary from synthetic cases, often tailored to quantum-specific issues, to established ones like MNIST~\citep{lecun2010mnist}, Fashion MNIST~\citep{xiao2017fashion}, Iris~\citep{anderson1936species,fisher1936use} and UCI\footnote{https://archive.ics.uci.edu/} datasets. The use of synthetic data in many studies suggests that QML is often operating in a proof-of-concept stage, possibly due to the NISQ constraints. We observed a recurring use of the MNIST dataset and its variants, which, while well-established in classical machine learning, raises questions in the quantum context. It’s well known that quantum advantage is not universal but problem-specific~\citep{ball2020physicists,herrmann2023quantum}. Therefore, the frequency of MNIST and other classical datasets might unintentionally misguide the field into a comparability trap with classical machine learning. This might hint at an ongoing struggle to balance between specificity and generalizability in QML models. The use of synthetic datasets, while necessary for proof-of-concept, should be complemented with real-world datasets to ensure the practicality of the models. Studies should ideally focus on problems that are inherently difficult for classical algorithms but are solvable more efficiently on a quantum setup. This observation leads to the question: Are we perhaps focusing on familiar grounds at the expense of uncovering quantum advantage?
 
On the other hand, the optimization of QML models is a complex task. The use of classical techniques such as Stochastic Gradient Descent (SGD) or backpropagation, and their quantum counterparts, is a topic of ongoing discussion between classical and quantum computation~\citep{lavrijsen2020classical,khairy2020learning}. However, these classical techniques may not be the most suitable for optimizing quantum circuits, particularly in the presence of quantum noise~\citep{khairy2020learning}. The optimization landscape of these models is highly non-convex, leading to low convergence rates for SGD. Moreover, the use of Nystrom approximation or Hamming distance-based optimization methods may be an attempt to circumvent quantum hardware limitations. Furthermore, the intrinsic difficulty of these optimization problems is highlighted by the NP-hard nature of training variational quantum algorithms~\citep{bittel2021training}. Even shallow variational quantum models, devoid of barren plateaus~\citep{marrero2021entanglement,arrasmith2021effect,wang2021noise,mcclean2018barren,holmes2022connecting,zhao2021analyzing}, have a superpolynomially small fraction of local minima within any constant energy from the global minimum~\citep{anschuetz2022quantum}. Just as in classical machine learning~\citep{bermeitinger2019singular,skorski2021revisiting}, these models become untrainable without an appropriate initial estimate of the optimal parameters. In addition, the exponential suppression of cost function differences in a barren plateau hampers the progress of gradient-free optimizers without exponential precision~\citep{arrasmith2021effect}. Learning can also be hindered without multiple copies of a state or if there is an excess of entanglement within the circuit~\citep{abbas2024quantum,marrero2021entanglement}. These challenges underscore the complexity of the optimization landscape in QML and the need for further research. While innovative, these methods should be critically assessed to ensure they do not compromise the potential advantages of a fully quantum approach, as they could result in quantum solutions that are neither faster nor more accurate than their classical counterparts~\citep{dunjko2018machine}.

\subsubsection{Algorithm performance} \label{subsubsec:performace}
\begin{table*}[htbp]
    \centering
    \caption{Prediction accuracy of various works on the MNIST and IRIS datasets. }
    \label{tab:datasetper}
    \begin{tabular}{|p{3.8cm}|p{1.5cm}|p{2.3cm}|p{1.3cm}|p{4.7cm}|}
    \toprule
    References & Dataset & number of classes & Accuracy in (\%) & Setting \\
    \midrule
    \cite{heyraud2022noisy} & MNIST & 3 - (3,6,8) & 94.50 & Decoherence (Spin Dephasing) \\ \hline %\midrule
    \cite{li2023ensemble} & MNIST & 4 - (1,3,5,7) & 95.00 & Noiseless \\ \hline %\hline
    \multirow{2}{*}{\cite{skolik2021layerwise}} &  \multirow{2}{*}{MNIST} & \multirow{2}{*}{ 2 - (6,9)} & 90.00 &Noiseless  \\ 
    % %\cline{4-5}%\cline{4-5} 
     & & & 73.00 & Shot Noise \\  \hline%\midrule
    \multirow{4}{*}{\cite{liang2021can}} & \multirow{4}{*}{MNIST} & \multirow{4}{*}{10} & 98.04 & Noiseless \\ %%\cline{4-5}
     &&& 88.24 & Flip Error(0.01) \\ 
     % %\cline{4-5}
     &&& 91.67 & Phase Error(0.01) \\ %%\cline{4-5}
     &&& 77.78 & Phase + Flip Error(0.01) \\ \hline %\midrule
    \multirow{2}{*}{\cite{huggins2019towards}} & \multirow{2}{*}{MNIST} & \multirow{2}{*}{2\footnotemark[1] -(4,9)} & 88.00 & Noiseless \\ %%\cline{4-5}
     &&& 80.60 & Amplitude(0.04) and dephasing(0.03) noise \\  \hline%\midrule
    \multirow{2}{*}{\cite{wang2022qoc}} & \multirow{2}{*}{MNIST} & 4-(0,1,2,3) & 63.70 & \multirow{2}{*}{Trained on-chip at ibmq\_jakarta}\\ %%\cline
    & & 2-(3,6) & 86.00 &  \\ \hline %\midrule
    \multirow{2}{*}{\cite{qian2022dilemma}} & \multirow{2}{*}{MNIST} & \multirow{2}{*}{10} & 94.00 & Noiseless \\ %%\cline{4-5}
    &&& 80.00 & Gate noise \\ \hline
    \cite{srikumar2022kernel} & FMNIST & 10 & 93.30 & Noiseless \\  \hline%\midrule
    \multirow{3}{*}{\cite{wang2021towards}} & \multirow{2}{*}{FMNIST} & \multirow{2}{*}{2 - (0,3)} & 96.00 & Noiseless  \\
     &&& 91.20 & Depolarizing rate(0.05) \\  \hline
    \multirow{2}{*}{\cite{wang2022qoc}} & \multirow{2}{*}{FMNIST} & 4 - (0,1,2,3) & 57.00 & Trained on ibmq\_manila \\%%\cline{3-5}
    && 2 - (3,6) & 90.70 & Trained on ibmq\_santiago \\  \hline%\midrule
    \cite{chen2021expressibility} & IRIS & 3 & 80.70 & Gate noise(0.01) \\  \hline%\midrule
    \cite{de2021quantum} & IRIS & 2-(0,1) & 98.89 & Noiseless \\ \hline
    \cite{abbas2021power} & IRIS & 2-(0-1) & $23.14$\footnotemark[2] & Trained on ibmq\_montreal \\ \hline
    \multirow{3}{*}{} & FMNIST& 2 &92.60  & \\
    \cite{canatar2022bandwidth}&KMNIST&2&91.50& Noiseless \\
    &PLAsTiCC&2&78.90& \\ \hline
    \cite{du2022problem} & FMNIST & 9 & 50.00 & Noiseless \\
    \botrule
    \end{tabular}  
    \footnote[1]{}{Section IV on \cite{huggins2019towards}  presents results for 45 cases $(10c2)$ combinations. Here, we only report the result that is severely impacted by the noise. \\
    \footnote[2]{}{It is training loss of a model as reported in \cite{abbas2021power}}
    }
\end{table*}
In this section, we present the performance of various models proposed among the selected papers. We focus the performance of these models on the classical dataset. A quick glimpse of these datasets is presented in Table~\ref{tab:datasetandOpt} and their performance in Table~\ref{tab:datasetper}.
We conveniently mapped the Fashion MNIST or FMINIST classes to \(0-9\) classes. When a study involves a subset of classes, it's indicated in the table under the \say{number of classes} column, e.g., (3 - (3,6,8)) represents a work focused on three specific classes—3, 6, and 8. We also note that accuracies are presented in percentages, and when individual noise rates were specified in the original works, we include them for context. 
It's apparent from Table~\ref{tab:datasetper} that the type and level of noise are pivotal in affecting the performance of the models. It is also evident that the performance of these models is sensitive to the presence of noise. For instance, the model by~\cite{liang2021can} shows an impressive $98.04\%$ accuracy on MNIST in a noiseless environment, but that number drops significantly under both flip error and phase error conditions. This sort of degradation is not unique and appears across multiple works, emphasizing the impact that different kinds of quantum noise can have on model performance. Similarly,~\cite{chen2021expressibility} model on the Iris dataset experiences an accuracy of $80.7\%$ under gate noise conditions, suggesting that even a relatively low noise level can have a measurable impact. The same pattern is observed in work by~\cite{wang2021towards} on the Fashion MNIST dataset, where the accuracy drops from 96\% in a noiseless setting to 91.2\% under a depolarizing rate of 0.05. Additionally, models trained on actual quantum hardware generally have lower accuracies compared to those trained in noiseless or simulated noisy environments.

This suggests that how a model is designed could be integral in mitigating specific noise types. More importantly, these shifts in accuracy due to the presence of noise highlight the challenges of operating in the NISQ era, mainly when the noise rates are non-negligible. Additionally, the performance accuracy of models trained on real quantum hardware raises questions about hardware-specific optimizations and the challenges this presents for reproducibility. The performance also seems to vary when only subsets of classes are considered, as seen in work like~\cite{wang2021towards} and~\cite{huggins2019towards}; this often raises concerns about the applicability of these models to real-world scenarios where class distributions are often imbalanced. It's also interesting to note that despite its simplicity, the Iris dataset tends to yield lower accuracies than more complex datasets like MNIST. Simpler datasets are often prone to be impacted by noise, and this is reflected in the performance of the models~\citep{khanal2023evaluating}. This suggests that the performance of QML models is not only sensitive to the type and level of noise but also to the complexity of the dataset.

\subsection{Bounds}\label{ssec:bounds}

In this section, we discuss different bounds and complexities proposed in the selected literature. This focus is to discuss the theoretical guarantees in QML proposed across literature under our inclusion criteria for model performance. The authors have proposed numerous bounds in different categories to provide a robust framework for evaluating the performance of QML models. One of the most prominent categories is the generalization bound, which is a metric central to any machine-learning task. We encourage readers to refer to the original works for a detailed derivation for each bound. In this paper, we discuss common properties observed across these bounds.

Generalization bound is an essential quantitative measure for assessing how well a model is expected to generalize to unseen data. In a QML setting, this bound often offers rich insights into the intricate interplay between quantum and classical computational resources beyond performance indicators. This contributes to our understanding of the capabilities and limitations of QML algorithms. A careful analysis of the generalization bounds provided in \cite{qi2023theoretical, caro2022generalization, huang_2021, liu2021rigorous, banchi2021generalization,caro2021encoding,abbas2021power,caro2023out,gibbs2024dynamical,gil2023understanding}—reveals a universal dependency on the dataset size $N$. This dependency aligns with well-established understandings in classical machine learning that larger datasets result in better model generalization~\citep{yasirlfdata}, reducing the model's uncertainty and error on unseen data. It's worth noting, however, that the influence of $N$ on the bounds isn't uniform across the board; the magnitude of its impact varies depending on other model parameters and method-specific assumptions. Furthermore, the generalization improvements appear to follow a sublinear trend, as most bounds show a $\mathcal{O}(\sqrt{N})$ behavior with respect to  $N$. Additionally, these bounds frequently incorporate specific model parameters—such as the Hilbert space dimension $d$ in \cite{qi2023theoretical,chen2021expressibility,huang_2021,caro2021encoding,abbas2021power,jerbi2023quantum}, the number of trainable quantum circuit gates $T$ in \cite{caro2022generalization,caro2023out}, and parameters $w$ in \cite{liu2021rigorous}. The bounds from~\cite{heyraud2022noisy,qi2023theoretical,caro2021encoding,abbas2021power,canatar2022bandwidth,gibbs2024dynamical} are additionally bounded in variables that are method specific. This implies two things: firstly, these bounds are often tailored to the specific algorithmic techniques or problem domains they are designed for, and secondly, the bounds suggest avenues for model fine-tuning, particularly by adjusting these specific parameters. Another noteworthy observation is that Quantum Kernel Theory is a recurring approach across many proposed generalization bounds~\citep{blank2020quantum,heyraud2022noisy,huang_2021,liu2021rigorous,wang2021towards,kubler2021inductive,canatar2022bandwidth}. In QML landscape, kernel theory appears to be serving as a foundational technique for constructing algorithms with both robust performance and theoretically justifiable generalization guarantees~\citep{schuld2019quantum,schuld2021supervised}. However, it is crucial to emphasize that the effectiveness of kernel methods trainability guarantees, due to convex loss landscapes~\citep{schuld2021supervised,schuld2022quantum}, hinges on the efficient estimation of kernel values to a sufficient precision~\citep{thanasilp2024exponential}. This is particularly challenging because, similar to the Barren plateau barrier in QNNs, hardware-induced noise in the near-term serves as a source of concentration for quantum kernel values to be exponentially concentrated towards some fixed value over different input data~\citep{thanasilp2024exponential}. 

On the other hand, Measurement complexity, defined as the number and nature of quantum measurements required to extract classical information from a quantum system, plays a crucial role in determining the generalization capabilities of QML models, particularly in noisy quantum systems. ~\cite{caro2021encoding} establishes a fundamental trade-off between the complexity of measurement observables and the amount of training data required for the effective generalization of QNNs. Their work demonstrates that while more complex measurements can enhance the expressivity of QNNs, they simultaneously demand larger training datasets to achieve robust generalization. This relationship is further explored in the context of quantum kernel methods, where~\cite{liu2021rigorous,gentinetta2024complexity} provide rigorous bounds on the number of measurement shots required to successfully train fidelity-based kernels~\citep{havlivcek2019supervised}. Furthermore,~\cite{wang2021towards} offers an optimistic perspective, showing that quantum kernel generalization can remain competitive with ideal scenarios when the number of measurements scales as $\mathcal{O}(N^3)$ provided the noise rate $p$ remains low. However,~\cite{thanasilp2024exponential} presents a contrasting view, demonstrating that under conditions of exponential concentration in quantum kernel values, the required number of measurement shots for precise kernel estimation scales exponentially. This dichotomy underscores the critical nature of measurement complexity in QML model performance, especially in noisy environments. Additionally, measurement complexity bounds proposed in~\cite{blank2020quantum,park2021robust} offer insights into the algorithm's resource requirements, specifically, the number of quantum measurements needed to maintain a certain level of accuracy in Kernel-based Quantum classifier models. These bounds exhibit a dependence on $p$. The bound \(\mathcal{O}\left( \frac{1}{(1-2p)^2}\right)\) from \cite{blank2020quantum}, valid for $p<0.5$, reveals a quadratic dependency on the noise rate $p$. This relationship implies that even small increases in noise can substantially elevate the measurement complexity, potentially limiting the algorithm's practicality in noisy environments. It is important to acknowledge that these bounds are primarily applicable within the specific context of kernel-based quantum models and may not necessarily extend to other QML frameworks. Regardless, collectively these works emphasize the delicate balance between measurement complexity, noise tolerance, and generalization performance in QML models.

Next, we discuss the query and runtime complexities, as these factors are instrumental in determining the practical feasibility of quantum approaches. Query complexity, which quantifies the number of interactions between an algorithm and an oracle or database, provides insight into the information-theoretic efficiency of quantum algorithms. \cite{du2018implementable} proposed a query complexity bound of \(\mathcal{O}(poly(\log(d)\sqrt{N}))\)  where \(d\) represents the feature space dimension, and \(N\) is the dataset size.  The bound exhibits a potential quantum advantage, as it scales polynomially with $\log(d)$ and only with the  $ \sqrt{N}$, potentially outperforming classical algorithms for high-dimensional data. However, this advantage must be weighed against the challenges posed by measurement complexity in noisy quantum systems, as discussed earlier. Runtime complexity, on the other hand, directly reflects the required computational time. The proposed runtime complexity bound \(\mathcal{O}(poly(\log d \log(d \log N)) \sqrt{\log N})\) by \cite{du2018implementable} demonstrates a more intricate scaling behavior, with polynomial dependencies on logarithmic terms of both $d$ and $N$. While this scaling is generally favorable compared to many classical algorithms, especially for large datasets, it is important to note that the actual performance advantage can be mitigated by the overheads associated with quantum state preparation and measurement, particularly in near-term devices. 

Furthermore, We observed that the authors also employ the Vapnik-Chervonenkis (VC) dimension to define generalization bounds. VC dimension plays a crucial role in understanding a model's capacity to generalize, providing an upper limit on the complexity of learnable functions. The VC dimension, defined as the largest number of points that a hypothesis class $\mathcal{H}$ can shatter (i.e., perfectly separate regardless of their labeling), takes on a distinctive form in quantum systems. ~\cite{chen2021expressibility} established a VC bound for quantum models: $2 \leq d_{vc} \leq (2\frac{n}{d} + 1)^{2d}$, where $d$ is the feature dimension and $n$ is the number of qubits. This bound highlights a fundamental difference from classical models: the VC dimension in quantum settings depends explicitly on the number of qubits, suggesting that the expressive power of quantum models scales with the size of the quantum system. Moreover, in noisy quantum environments, the VC bound incorporates an additional dependency on the circuit depth $L_c$, as observed in works by~\cite{caro2021encoding,abbas2021power,caro2023out,caro2022generalization}. This three-way dependency on feature dimension, qubit count, and circuit depth represents a significant departure from classical machine learning, where generalization typically depends primarily on the feature space dimension and sample size. The inclusion of circuit depth in the VC bound for noisy quantum systems emphasizes the intricate relationship between model complexity, noise, and generalization in QML. It suggests that deeper quantum circuits, while potentially more expressive, may face greater challenges in generalization, especially in the presence of noise. Furthermore, these bounds hint at both potential advantages and challenges for QML models. On one hand, the dependence on qubit count suggests that quantum models might offer enhanced expressive power that scales efficiently with system size. On the other hand, the sensitivity to circuit depth in noisy settings underscores the challenges of maintaining this expressivity in practical, noisy quantum devices. 

\subsection{Computing Platforms}\label{ssec:platforms}
\begin{table*}[htbp]
    \centering
    \caption{Various quantum computing platforms used for experiments in literature for QML experiments}
    \label{tab:platforms}
    \begin{adjustbox}{width=\textwidth,totalheight=\textheight,keepaspectratio}
    \begin{tabular}{ p{3.5cm}| p{0.7cm} p{1.6cm} p{0.9cm} p{1cm} p{1.4cm} p{1cm} c p{1cm} p{1cm}}
    Reference & Qiskit & IBM Quantum Platform\footnotemark[1] & PyQuil & GENCI & Pennylane & Julia & Tensorflow & MGCF & QuTip \\
    \hline
    \cite{basu2022qer} &  \checkmark  & \checkmark & & & & & &   \\
    \cite{du2018implementable} & & & \checkmark & & & & & &  \\
    \cite{blank2020quantum} &\checkmark & \checkmark & & & && & &   \\
    \cite{chen2021expressibility} & \checkmark & & & & & & &  & \\
    \cite{srikumar2022kernel} & \checkmark & \checkmark & & & && & &   \\
    \cite{heyraud2022noisy} & & & & \checkmark & & & & &  \\
    \cite{zardini2022implementation} & \checkmark & & & && & & &   \\
    \cite{qi2023theoretical} & & & & & \checkmark & & &  & \\
    \cite{li2023ensemble} & & & & & & \checkmark & &   &\\
    \cite{huang_2021} & & & & & & & \checkmark &  &\\
    \cite{park2021robust} & &  \checkmark  & & & & &&   \\
    \cite{skolik2021layerwise} & & & & & & & \checkmark &&   \\
    \cite{liang2021can} & \checkmark & \checkmark & & & & && &   \\
    \cite{wang2021towards} & \checkmark & \checkmark & & & && & &   \\
    \cite{coyle2020born} & & & \checkmark  & & & & &   & \\
    \cite{hamilton2020error} & & \checkmark & & & & & & &  \\
    \cite{huggins2019towards} & & & & & & & &   \checkmark& \\
    \cite{wang2022qoc} & & \checkmark & & & & & &   & \\
    \cite{kim2020quantum} & \checkmark & \checkmark &&&&&&\\
    \cite{de2021quantum}& \checkmark & \checkmark &&&&&& \\
    \cite{qian2022dilemma} & & \checkmark &&& \checkmark& &&\\
    \cite{wang2021noise} && \checkmark &&&&&&\\
    \cite{abbas2021power} & & \checkmark &&&&&& \\ 
    \cite{kubler2021inductive}& &&&&\checkmark&&&& \\
    \cite{canatar2022bandwidth} & \checkmark&&&&&&&& \\
    \cite{jerbi2023quantum}& & & & & & & \checkmark& &  \\
    \cite{caro2023out}&&\checkmark&&&&&&& \\
    \cite{gibbs2024dynamical}&&\checkmark&&&&&&& \\
    \cite{haug2023generalization}&&&&&&&&&\checkmark \\
    \cite{gil2023understanding}&&&&&&&\checkmark && \\
    \hline
    \end{tabular}
    \end{adjustbox}
    \footnote[1]{}{Refers to multiple IBM Quantum devices. Details on these platforms are provided in section~\ref{ssec:platforms}.}
    
\end{table*}
% Computing Platforms
The choice of quantum computing platforms is a crucial aspect of QML research, as it directly impacts the experimental feasibility and the generalizability of the results. In Table~\ref{tab:platforms}, we provide the list of quantum computing platforms used for an experiment by the work listed in Table~\ref{tab:finalPapers}. The IBM Quantum Platform, such as Melbourne, Ourense, Rome, and Montreal, appears to be quite popular. The Melbourne processor, a \(15\) qubits system retired on  $08/09/2021$, tends to be the most utilized in the IBM quantum series, likely due to its higher qubit count and may be due to its early market entry. Specifically, Melbourne is used in works by \cite{basu2022qer, srikumar2022kernel, wang2021noise, kim2020quantum, de2021quantum}, Ourense by \cite{blank2020quantum, wang2021towards}, Rome by \cite{park2021robust}, and Montreal by \cite{wang2021noise,liang2021can,abbas2021power}. While using these platforms is understandable, given their accessibility and the extensive support provided by IBM, it's important to note that the choice of platform can impact the results and potentially create research bias. For instance, the noise rates and error models of these platforms can vary, leading to different performance outcomes for the same model. 

Another crucial observation is the use of Qiskit, a popular quantum computing software development kit (SDK) by IBM. Qiskit is used in works by \cite{basu2022qer, blank2020quantum, chen2021expressibility, srikumar2022kernel, zardini2022implementation, liang2021can, kim2020quantum, de2021quantum,canatar2022bandwidth}. This is not surprising, given that Qiskit is one of the most widely used quantum computing frameworks. Software simulation often provides the first line of feasibility tests for quantum algorithms. However, these results can be optimistic compared to the NISQ devices since the simulators usually lack noise models under normal settings. Furthermore, the use of Pennylane only in \citep{qi2023theoretical,qian2022dilemma,kubler2021inductive} is intriguing, mainly because Pennylane's focus on differentiable quantum computing makes it particularly well-suited for hybrid quantum-classical models. Additionally, the use of Julia in \cite{li2023ensemble} and TensorFlow Quantum in \cite{huang_2021, skolik2021layerwise} might hint at a move towards using more traditional machine learning frameworks. However, one could argue that these choices should not merely be about convenience or the ease of integration with classical models. They should also be critically evaluated for their capability to handle quantum-specific issues, such as error correction or the intricacies of quantum gate operations.

It is imperative to acknowledge, however, that this distribution does not necessarily mirror the broader QML community's platform preferences. Preliminary observations and informal surveys within the community suggest a significant and possibly growing interest in platforms like PennyLane, Tensorflow quantum, and QuTip, which may not be fully represented in our dataset.~\citep{kordzanganeh2023benchmarking,hevia2022quantumpath,serrano2022quantum} are some of the works that provide a comprehensive study on quantum computing platforms used in literature. 

In our findings, the platform selections are far from arbitrary, influenced by factors such as ease of use, capability, and perhaps even academic and commercial affiliations. Observing these diverse platforms used in the experiment raises the issue of reproducibility. How many of these papers provide adequate information for replicating their experiments on other platforms?\footnote{Note that Pennylane has various plugins for an accessible dispatch of quantum functions to different quantum devices.} The field risks becoming fragmented if results obtained on one platform cannot be compared or reproduced on another.

\subsection{Research Approach}\label{ssec:researchapproach}
Among the selected papers, we find that the research approaches are diverse, with a mix of theoretical and empirical work. Most of the research appears to use a theoretical approach with empirical validation. Such an approach is necessary, especially in the NISQ era, where empirical work can offer immediate insights into error rates, robustness, and other practical considerations that are crucial for theoretical generalizations. It is interesting to note that various articles address core concerns in machine learning from a quantum perspective, such as generalization bound~\citep{,abbas2021power,canatar2022bandwidth,caro2023out,gibbs2024dynamical,gil2023understanding,banchi2021generalization,caro2022generalization,huang_2021,wang2021towards,chen2021expressibility}, kernel methods~\citep{heyraud2022noisy,wang2021towards,srikumar2022kernel,blank2020quantum,canatar2022bandwidth,kubler2021inductive,jerbi2023quantum}\footnote{\cite{jerbi2023quantum} provides a samples complexity which can be converted in to generalization bound in most of the cases}, and ensemble learning~\citep{basu2022qer,srikumar2022kernel,li2023ensemble,caro2023out}. Unlike other works discussed in this review,~\cite {jerbi2023quantum} provides the lower bound for qubit complexity for QNNs in an ideal setting. It is no surprise that these bounds are expressed in terms of the feature space dimension $d$. This suggests that the field is actively working towards addressing foundational learning problems, such as the true capabilities and limitations of QML models and their robustness and error tolerance in a NISQ environment. However, given that many of these approaches combine theory and experiment, one can hypothesize that the field is still working towards addressing foundational learning problems, such as the true capabilities and limitations of QML models and their robustness and error tolerance in a NISQ environment. We provide such empirical results on classical data in Table~\ref{tab:datasetper}.

Another interesting observation is the use of quantum kernel methods. Kernel methods seem to be an attempt to utilize classical machine learning techniques in quantum architectures to benefit from the mathematical rigor of kernel theory while aiming to harness quantum advantages. This could offer better generalization for QML models~\cite{schuld2021machine}. Recent research has been motivated towards the quantum kernel-based theoretical and experimental advancement, with studies establishing a connection between supervised learning and quantum kernels~\cite{schuld2019quantum}. For a class of machine learning problems, quantum kernel methods can solve them efficiently, which is hard for all classical methods~\citep{liu2021rigorous}. Furthermore, expressivity and generalization capacity of quantum kernels have been investigated in studies by~\cite{chen2021expressibility,heyraud2022noisy}. \cite{huang_2021} found that data availability can modulate the computational hardness of learning tasks. However,~\cite{heyraud2022noisy,wang2021towards} reveal that the inner products or fidelity measures constituting quantum kernels can be particularly susceptible to noise, thereby affecting their overall performance and the reliability of the QML models built upon them. 

Nonetheless, while quantum kernels have the potential to be advantageous in NISQ settings due to their ability to find better or equally good quantum models compared to variational circuit training~\citep{schuld2021supervised}, it is important to acknowledge that they might suffer from exponential concentration under certain conditions~\citep{thanasilp2024exponential}. This phenomenon, while not necessarily affecting the trainability of quantum kernels, can lead to poor generalization, where the model's predictions on unseen data become independent of the input data, thereby undermining the expected advantages. ~\citep{thanasilp2024exponential} identified the expressivity of data embedding, global measurements, entanglement, and noise as a source of the concentration in quantum kernel models. This trade-off between the optimization advantages of quantum kernels and their potential generalization challenges is a careful consideration for kernel model design and implementation. 

Furthermore, the efficiency of data encoding plays a crucial role, with compact encoding schemes potentially minimizing the number of gates required~\citep{schuld2021effect,gan2023unified}. While both quantum kernel methods and QNNs utilize quantum circuits to encode data into quantum states, the role and function of these circuits differ between the two approaches. In quantum kernel methods, the quantum feature map is explicitly designed to project data into a high-dimensional quantum space, where classical algorithms may then operate more effectively, depending on the problem. This can result in simpler classification circuits, as the complexity is handled by the classical algorithm post-quantum feature mapping. In contrast, QNNs embed the data encoding within a parameterized quantum circuit, where the entire model, including the quantum feature map, is optimized during training. As a result, the simplification observed in quantum kernel methods may not directly apply to QNNs. This potential for shallower circuits and fewer gates in quantum kernel methods aligns well with the limitations of NISQ devices, where circuit depth is constrained by noise~\citep{wang2021towards}. However, it's important to note that the best choice between quantum kernels and other QML methods depends on the specific dataset, data-encoding ansatz, learning problem, and the available hardware~\citep{jerbi2023quantum}. On the other hand, QNNs can offer significant expressive power and flexibility in learning complex patterns~\citep{abbas2021power}. However, this expressivity frequently comes with increased computation resource requirements, especially with deeper architectures~\citep{skolik2021layerwise,qian2022dilemma}. This can pose challenges for NISQ devices. Furthermore, concerns exist regarding the learnability and trainability of QNNs in noisy settings~\citep{qian2022dilemma,du2021learnability}. In contrast, quantum kernel methods often benefit from guarantees of convex optimization and potential advantages in resource utilization~\citep{park2021robust} but suffer from exponential concentration~\citep{thanasilp2024exponential}. Yet, their data scaling limitations and reliance on specific kernel functions can restrict their applicability. Additionally, data requirements for QNNs often scale with circuit depth and problem complexity, potentially becoming substantial~\citep{skolik2021layerwise}.

Another research approach that appears in the final paper list is ensemble learning for error mitigation. \citep{li2023ensemble,srikumar2022kernel} used this approach for error mitigation, suggesting that it might be a practical strategy to make quantum algorithms more robust. However, ensemble methods inherently require the collection of multiple models, which could be resource-intensive. We do not find any work explicitly addressing the trade-off between improved performance and increased resource utilization.

\section{Discussion}\label{sec:discussion}
In this section, we discuss the implications of our findings and the limitations of our methodology. We discuss the interplay between generalization bounds and others' complexity, the pitfalls in a dataset and optimization choices, platform standardization vs. research bias, the theoretical-experimental divide, fragmented research approaches, and the implications for the NISQ era. Our analysis suggests that there is a gap in the techniques and tendencies that could advance QML research in certain directions, both limiting and enabling.
% \subsection{Interplay between Generalization Bounds and Measurement Complexity}\label{ssec:interplay}

The generalization error bound is interlinked with the measurement, sample, and qubits complexity. While generalization bounds give a theoretical metric to evaluate model robustness, qubits and sample complexity provide a practical measure of the resources required to achieve this robustness, their practical utility depends on the feasibility of measurement. It's striking that all these metrics are yet to be optimized together in the literature, an oversight that could hinder the transition from theory to practice. What remains unavailable is a framework that can combine these aspects, allowing for not just error prediction but also its empirical verification on actual quantum hardware.

% \subsection{Pitfalls in Dataset and Optimization Choices}\label{ssec:pitfalls}

The use of classical datasets like MNIST and IRIS raises the concern of what we term as \say{familiarity bias}. This leads to inevitable comparisons with classical algorithms, obscuring the specific advantages that quantum algorithms may bring. Concurrently, the dataset choices show a clear division between synthetic datasets and real-world datasets, with each having its own set of advantages and disadvantages. On one side, synthetic datasets may allow for a deeper understanding of quantum-specific phenomena but at the expense of broader applicability. On the other hand, classical datasets may not necessarily help demonstrate the quantum advantage. 

% \subsection{Platform Standardization vs. Research Bias}\label{ssec:platformbias}

The frequent use of IBM's quantum platforms suggests a potential trend toward platform standardization, but it also raises a concern about research bias. While standardization can facilitate result comparison and replication, the field should be wary of a one-size-fits-all approach. Different platforms have varying noise models, gate fidelities, and connectivity architectures that can significantly impact algorithm performance and, hence, the generalizability of the research findings.

% \subsection{Theoretical-Experimental Divide}\label{ssec:theoryexpdivide}

The majority of theoretical work with empirical validation indicates a field that is cautiously optimistic. However, it's essential to question whether the empirical work truly validates the theory or simply provides a proof of concept under idealized/simulated conditions. The field needs to bridge the gap between theory and practice, ensuring that theoretical insights are validated in realistic settings. 

% \subsection{Fragmented Research Approaches}\label{ssec:fragmentation}

Furthermore, the evident focus on kernel methods indicates a field searching for a stable theoretical foundation. Kernel methods offer mathematical rigor but could be susceptible to noise, whereas ensemble learning shows promise in error mitigation but may raise questions on resource optimization. These diverging approaches risk fragmenting the field unless they are part of a more extensive, unified strategy.

% \subsection{Implications for the NISQ Era}\label{ssec:nisqimplications}
The QML community has yet to fully adapt to a unique set of constraints imposed by the NISQ era. While work on generalization bounds and other complexities indirectly acknowledges these limitations, direct strategies to navigate the NISQ landscape are conspicuously absent. The modifications or \say{quantum patches} to classical techniques (like SGD) can be seen as stop-gap measures but hardly a long-term solution.

% \subsection{Future Outlook}\label{ssec:futureoutlook}
Our analysis suggests a more unified and collaborative approach. This could involve sharing across generalization bounds, measurement complexity, dataset selection, optimization techniques, and platform-agnostic strategies. \cite{caro2023information} proposed to establish a general mathematical framework for quantum learning. However, it is essential to recognize that the diversity of machine learning models is a strength, not a weakness. Each model offers unique benefits and limitations, and there is no one-size-fits-all solution. The right model depends on the problem at hand.

 %subsection{Methodology and Limitations}\label{ssec:methodology}
The methodology of this survey is based on a systematic literature review, which is a well-established method for synthesizing and analyzing information from previously published literature. However, it's important to highlight that our search and filtration process is subject to certain constraints. The scope of our search was limited to certain academic databases and English-language publications, which may have inadvertently excluded relevant international research, details on section~\ref{subsec:planning},\ref{subsec:IncExc}. Furthermore, the database search was conducted in early $2023$ to collect the relevant papers from $2010$ to $2023$\footnote{Some of the latest papers were hand-picked as the snowballing process.} timeframe. The field of QML is rapidly evolving. Therefore, it is possible that some recent research may not have been included in our analysis. Furthermore, our inclusion and exclusion criteria were designed to maintain focus and relevance, but they may also introduce a degree of selection bias. By recognizing these limitations, we aim to ensure that the findings of our review are interpreted with an appropriate level of scrutiny and consideration for the broader research landscape.

\section{Conclusion}\label{sec:conclusion}
This survey offered a comprehensive analysis of the current state of supervised QML, focusing on generalization bounds, measurement complexities, datasets, optimization techniques, computing platforms, and research approaches. Our findings reveal a field's formative stages, struggling to balance theoretical robustness and practical applicability.  

Generalization bounds are a pivotal element in evaluating QML models, revealing a dependency on dataset size, feature space dimension, and model-specific parameters. These bounds have also been closely tied to measurement sample and qubits complexities, other vital aspects of assessing the practicality of QML, especially in a NISQ environment. In practice, the descent number of research work~\citep{heyraud2022noisy,liang2021can,huggins2019towards,wang2022qoc,canatar2022bandwidth,wang2021towards} shows a trend toward using well-known classical datasets and optimization techniques. We presented the performance metrics of these works on IRIS, MNIST, and FMNIST datasets with gradient descent being the favorite choice for optimization. This focus raises questions about the field's ability to tackle quantum-specific challenges, calling for broader dataset and algorithmic diversity. It is worth noting that the use of classical datasets and optimization techniques is not necessarily a limitation but rather an intermediate step to understanding the possibilities and limitations of QML in the NISQ era. However, it is important to mention that there has been a shift to develop GEB in quantum setting with an experiment on quantum data~\citep{caro2021encoding,caro2023out,gibbs2024dynamical,gil2023understanding,haug2023generalization}. Despite the diverse directions of quantum learning theory,~\cite{caro2023information} has propose a formalism for describing quantum learning tasks that involve training on classical-quantum data and then testing the learned hypothesis on new data. These new directions show that the QML community is moving towards more quantum-specific approaches.

Likewise, the frequent use of specific platforms, particularly from IBM, hints at a potential research bias and raises questions about cross-platform reproducibility. The research approaches reflect a blend of theoretical and empirical work, with a focus on foundational problems in machine learning. Prominent techniques include kernel methods, ensemble learning, and randomized circuit learning. Yet, several unanswered questions remain, especially concerning the trade-offs between performance and resource utilization. \cite{du2022problem} has identified that the multi-class classification power of QNN classifiers is dominated by the training loss rather than generalization ability. ~\cite{gil2023understanding} have challenged current understandings of generalization in QML. They presented a compelling result that showed that state-of-the-art QNNs can memorize random data, defying standard theories about how these models generalize. This questions the ability of current frameworks to guarantee how well a model performs on new data. Their work is key to understanding and making better QML algorithms. 

The field of QML presents both opportunities and challenges. This survey focused on generalization bounds for supervised QML, but the field is actively exploring other branches, such as model expressibility and trainability. \cite{du2022efficient} analyzed the expressivity of VQCs using qudits, finding an upper bound determined by the number of quantum gates and measurement observables used in the ansatz. Further research into qudit-based quantum computation and the expressibility and training of QML models can be promising future directions for the field.

As we navigate the intricacies of quantum computing, the task ahead is to address the gaps identified in this survey, striving for methodological diversity, practical applicability, and cross-platform standardization.
We have identified several avenues for future work:
\begin{itemize}
    \item A closer investigation of the trade-offs between generalization and computational costs.
    \item The need for more diverse datasets tailored for quantum phenomena.
    \item Scrutiny of the impact of platform biases on research outcomes.
    \item Evaluation of optimization techniques specific to quantum systems.
    \item Exploration of resource utilization in ensemble learning and other error mitigation strategies.
    \item A comprehensive theoretical development to lay the foundation for the unified understanding of quantum kernels~\cite{gan2023unified} and quantum learning~\citep{caro2023information}.
\end{itemize}

By confronting these challenges, the QML community can move closer to realizing the full potential of quantum computing in machine learning applications. 

% \bmhead{Acknowledgments}
% The authors want to thank the reviewers for their thoughtful feedback and suggestions during the review process.

\section*{Declarations}
\bmhead{Funding} 
Part of this work was performed while P.R. was funded by the National Science Foundation under Grant Nos. 2136961, and 2210091, and 1905043. 
The views expressed herein are solely those of the author(s) and do not necessarily reflect those of the National Science Foundation.

\bmhead{Conflict of interest}
The authors have no relevant financial or non-financial interests to disclose.

\bmhead{Availability of data and materials}
Data sharing does not apply to this article as no datasets were generated or analyzed during the current study. This article is a systematic literature review, and as such, it synthesizes and analyzes information from previously published literature.

\bibliography{ref}

\end{document}